%% file: main.tex
\documentclass[conference]{IEEEtran}  
\IEEEoverridecommandlockouts

\usepackage{graphicx} 
\usepackage{gensymb}
\usepackage{subcaption}
\usepackage[acronym,nonumberlist]{glossaries}
\usepackage[utf8]{inputenc}
\usepackage[T1]{fontenc}
\usepackage{lmodern}
\usepackage{microtype}
\usepackage{enumitem}
\usepackage[letterpaper, left=0.65in, right=0.65in, top=0.75in, bottom=1in]{geometry}
\usepackage[colorlinks=false, pdfborder={0 0 0}, hidelinks]{hyperref}
\hypersetup{draft} 

\makeglossaries
\usepackage[english]{babel}

\usepackage{comment}
\usepackage{amsfonts}
\usepackage{bbm}
\usepackage{caption}
\usepackage{graphicx}
\usepackage{multirow}
\usepackage{tikz}

\usepackage{amsmath}
\usepackage{amsthm}
\newtheorem{theorem}{Theorem}
\newtheorem{definition}{Definition}

\usepackage[ruled,vlined]{algorithm2e}

\usepackage{cite}

\usepackage{booktabs}

\usepackage{tikz}
\usepackage{xcolor}\usetikzlibrary{shapes.geometric, arrows, automata, positioning}
\def\BibTeX{{\rm B\kern-.05em{\sc i\kern-.025em b}\kern-.08em
    T\kern-.1667em\lower.7ex\hbox{E}\kern-.125emX}}

\title{Adaptive Scheduling: A Reinforcement Learning Whittle Index Approach for Wireless Sensor Networks}

\author{
\IEEEauthorblockN{
Sokipriala Jonah\IEEEauthorrefmark{1},
Seong Ki Yoo\IEEEauthorrefmark{2},
Saurav Sthapit\IEEEauthorrefmark{3}
}
\IEEEauthorblockA{\IEEEauthorrefmark{1}
Centre for Computational Science and Mathematical Modelling, Coventry University, Coventry, UK}
\IEEEauthorblockA{\IEEEauthorrefmark{2}
Wireless Transmission, BT Group, Ipswich, UK}
\IEEEauthorblockA{\IEEEauthorrefmark{3}
Software and Security Group, Analog Devices, Edinburgh, UK}
}

\date{March 2025}

\begin{document}
\input{acronyms_list}

\input{glossaries_list}
\maketitle

\begin{abstract}
We propose a reinforcement learning--based scheduling framework for Restless Multi-Armed Bandit (RMAB) problems, centred on a Whittle Index Q-Learning policy with Upper Confidence Bound (\gls{wiql}-\gls{ucb}) exploration. Unlike existing approaches that rely on fixed or adaptive $\epsilon$-greedy strategies and require careful hyperparameter tuning, the proposed method eliminates problem-specific tuning and is therefore more generalisable across diverse RMAB settings. We evaluate \gls{wiql}-\gls{ucb} on standard RMAB benchmarks and on a practical sensor scheduling application based on the Age of Incorrect Information (AoII), using an edge-based estimation scheme that requires no prior knowledge of system dynamics.
Experimental results demonstrate that \gls{wiql}-\gls{ucb} achieves near-optimal performance while significantly improving efficiency. At a representative problem size (\(N=15, M=3\)), the proposed method requires only about 600~bytes of memory, compared to several kilobytes for tabular Q-learning and hundreds of kilobytes to megabytes for deep reinforcement learning baselines. In addition, \gls{wiql}-\gls{ucb} attains sub-millisecond per-decision runtimes and is several times faster than deep RL approaches, while maintaining competitive performance.
These results demonstrate that \gls{wiql}-\gls{ucb} consistently outperforms both non–Whittle-based and Whittle-index learning baselines across diverse RMAB settings.

\end{abstract}

\begin{IEEEkeywords}
Wireless sensor networks, Age of Incorrect Information, Reinforcement learning, Restless Multi-Armed Bandit, Upper Confidence Bound
\end{IEEEkeywords}
\smallskip
\noindent\textit{Version note:}
This manuscript is a substantially extended version of our earlier arXiv preprint \cite{jonah2026reinforcement}. The present version introduces a UCB-based Whittle index learning framework and includes a significantly expanded experimental evaluation across multiple RMAB problems, with systematic comparisons against both Whittle-index-based and non-Whittle-index policies.
\bigskip

\section{Introduction}
\label{sec:introduction}
In many remote monitoring applications, various sensors are deployed to track environmental conditions in applications such as smart homes, agricultural, and industrial monitoring. As sensor nodes are resource-constrained and typically battery-powered, continuous data collection and packet transmission consume significant amounts of energy. Additionally, frequent and indiscriminate transmissions increase storage requirements at the remote monitoring system without necessarily improving the accuracy of the state estimation. These challenges have led to the adoption of goal oriented scheduling techniques, which prioritise transmissions based on the value of the information they provide or the goal of the remote monitoring system \cite{zhang2023goal,jonah2025comparison}. 

In a standard application, the network objective is to transmit packets with high reliability from the transmitter to the receiver. For instance, consider a simple monitoring scenario where the goal is to keep the room temperature warm. In a traditional approach, all temperature readings would typically be transmitted. However, in a goal-oriented system, data is transmitted only when there is a change in the current state of the room.
In this case, processing can be performed at the edge, on an \gls{iot} or \gls{wsns} node where the data is generated. This enables the transmission of only information that is relevant to the underlying control or monitoring objective. Transmitting all raw sensing data in such scenarios is often highly inefficient and unnecessary, particularly in bandwidth- and energy-constrained industrial deployments.
In \cite{gaura2013edge}, an edge-mining approach termed Class-Act is proposed, in which accelerometer data are transmitted only when a change in activity state is detected, such as transitions between standing, sitting, or walking.

By suppressing redundant transmissions during periods of unchanged behaviour, the approach significantly reduces communication overhead while preserving task-relevant information.
Related goal-driven communication principles have also been applied to real-time tracking and control of autonomous systems, where transmissions are triggered only when necessary to maintain accurate state estimation or control performance \cite{pappas2021goal}.  
Similarly, scheduling based on the value of information in wireless sensing systems has been investigated in \cite{bidoki2018joint,jonah2025energy}.  
These approaches prioritise task relevance over raw data fidelity and demonstrate that selective, goal-oriented information transmission can significantly improve efficiency under stringent communication constraints.

A key challenge in this problem is that the sink (or gateway), which can only poll a subset of the  sensors at each time step due to channel constraints, does not have perfect knowledge of the monitored process at the nodes. To address this, we propose a reinforcement learning-based polling strategy that selects the sensors expected to provide the most valuable updates, thereby optimising state estimation at the sink. We build upon the edge mining approach used in \cite{gaura2013edge}, where raw sensor data is processed at the node side before transmission, ensuring that only meaningful information which improves the state estimation of the node at the sink side is sent to the sink.

Building on this, we employ semantics-aware \gls{aoii}, which quantifies how outdated the state estimate at the sink is and serves as a metric for prioritising which sensors to poll.
To solve this scheduling problem, we develop a reinforcement learning solution to learn the Whittle index, a technique that has been shown to yield near-optimal performance in scheduling problems where the transition probabilities of the underlying system are known. However, even under known transition probabilities, computing an optimal scheduling policy for selecting \( M \) out of \( N \) processes in systems with large state spaces remains computationally expensive.  

In contrast to previous works discussed in Section \ref{section:related_work}, which assume known transition probabilities or complete knowledge of the system state at the sink, our approach makes no such assumptions. Although this increases the complexity of obtaining an optimal solution, it better reflects real-world sensor monitoring scenarios where state transitions are typically unknown. Consequently, we adopt the edge mining approach from \cite{gaura2013edge} to directly estimate \gls{aoii}. We use an online reinforcement learnin based approach to learn an optimal scheduling policy.
Contrary to previous work, we conduct extensive experiments using synthetic data that represents typical sensor monitoring and control applications to evaluate the performance of our approach. 
We further validate the generalisability of the technique across diverse domains, including maintenance scheduling and preventive healthcare applications.
Specifically, we propose a \gls{ucb}-based Q-learning method to learn the \gls{wiql} policy, which does not require hyperparameter tuning in contrast to works in \cite{fu2019towards,avrachenkov2022whittle}. We compare the performance of our \gls{wiql} method with other Q-learning based Whittle index policies and state-of-the-art adaptive scheduling algorithms across various scenarios.
The primary contributions of this work are as follows:  
\begin{itemize}
    \item We present a practical technique for accurately estimating the \gls{aoii} at the sink without relying on assumptions about the underlying node dynamics. This is accomplished using an edge mining approach that enables local processing at the data source.
    
    \item We introduce \gls{wiql}-\gls{ucb}, an asymptotically optimal learning algorithm that eliminates the need for manual hyperparameter tuning and achieves lower average \gls{aoii} compared to existing techniques.
    
    \item We demonstrate that \gls{wiql}-\gls{ucb} generalises effectively across diverse application domains, outperforming state-of-the-art methods for online learning of the Whittle index.
\end{itemize}

\section{Background and Related Works}
\label{section:related_work}

In a standard sensor monitoring process, particularly in less dense networks, packets are transmitted either at regular time intervals or in an event-based manner, depending on the occurrence of an event, usually in a distributed fashion. However, in large-scale sensor networks or Internet of Things (IoT) systems, as the network size increases, channel constraints lead to collisions, resulting in packet failures and retransmissions. To mitigate these issues, various scheduling techniques have been proposed where the gateway, sink, or base station schedules transmissions.

One widely studied approach in the literature is the Kalman filter estimation technique, where scheduling decisions are made based on minimizing the estimation covariance trace \cite{ajit2023energy,han2016opportunistic,liu2021remote,li2020application,wei2022multisensor}. Additionally, reinforcement learning techniques have been applied to improve Kalman filter-based scheduling methods \cite{leong2020deep,demirel2018deepcas,alali2024deep,yang2022optimal}. In these techniques, scheduling decisions are made to minimise the covariance trace, thereby effectively reducing the \gls{mse} of the estimator. These systems typically assume that the gateway has full knowledge of the underlying dynamics of the processes being monitored by the sensor nodes. 
Additionally, recent research in goal-oriented and semantics-aware communication has shown that minimising the \gls{mse} in sensor estimation does not necessarily align with optimising the actual goal of the sensor monitoring system~\cite{maatouk2022age,kriouile2021minimizing}.

\subsection{Age of incorrect information}

More recently, goal oriented and semantic aware scheduling techniques such as the \gls{aoi} have been proposed, where the scheduler polls nodes based on the timestamp of the last successfully received update \cite{saurav2023scheduling,jin2022deep,jhunjhunwala2018age,corneo2019age}.
Specifically, \gls{aoi} captures the time elapsed since the most recent successful update was received at the monitor, increasing linearly with time until a new update arrives~\cite{liu2022wireless,jin2022deep,li2020age}. Formally, it is defined as:
\begin{equation}
    \delta \text{AoI}(t) = t - g(t)
\end{equation}
where \(g(t)\) denotes the time at which the last successful update was received at the monitor.
However, several studies have demonstrated that scheduling based solely on \gls{aoi} does not necessarily improve the \gls{voi} at the sink \cite{talli2024push,maatouk2022age,jonah2025energy}.
Frequent transmissions that rely solely on the last transmission attempt without assessing whether there has been a change in the environment or the observed phenomenon do not add value to the information already available at the sink, particularly when the environment remains stable since the last update. Moreover, such transmissions consume energy, which is a critical design consideration in battery powered, resource constrained devices.

Another semantic-aware metric gaining attention is the \gls{aoii}, which integrates time-based and estimation error metrics. This metric prioritises updates based on the expected change in the environment, ensuring both timely updates and accurate state estimation at the sink \cite{maatouk2022age,gunduz2022beyond}.
The \gls{aoii} quantifies how outdated the observation at the remote observer is and used to ensure nodes which when scheduled in a channel constrained network improve the the \gls{aoii} are scheduled to reduce the average \gls{aoii}.
Minimising the penalty \gls{aoii} ensures that updates are not only timely but also relevant prioritising the delivery of state changes that matter. This makes AoII a more robust metric for applications where update value, rather than frequency alone, is essential. The \gls{aoii} penalty function is defined as:

\begin{equation}
     {\text{AoII}}(t) = f(t) \times g(x(t), \hat{x}(t))
\end{equation}
\begin{itemize}
    \item \(f(t)\): A time-dependent penalty function that grows with time and
    represents the cost of not being aware of the correct status of the process at the last updated time. 
    \item \( g(x(t), \hat{x}(t)) \): A function that represents the error-based penalty (e.g, indicator, squared error, threshold error) details can be found in \cite{maatouk2022age}. 
    the error is the difference between the actual state of the process \(x(t)\) and the current estimate \(\hat{x}(t)\) at the monitor.
\end{itemize}
The \gls{aoii} has been shown to evolve as a \gls{rmab} in recent works \cite{maatouk2022age,ayik2023optimization}, where each sensor node can be modelled as a decision process whose state continues to evolve even when it is not actively scheduled. This structure naturally leads to a \gls{rmab} formulation.

\subsection{Restless Multi-Armed Bandits}
\label{sec:rmab}

An \gls{rmab} extends the classical Multi-Armed Bandit problem by allowing all arms to evolve according to individual \gls{mdp}s, regardless of whether they are selected. In contrast, traditional bandit and \gls{mdp} formulations assume that only the chosen process (or arm) evolves, while the states of inactive arms remain frozen. This persistent evolution across all arms introduces significant computational complexity, as the joint state space grows exponentially with the number of arms, rendering exact optimisation intractable in most practical settings.

At each decision epoch, the scheduler is constrained to select a subset of \(M\) arms from a total of \(N\), which introduces a combinatorial action space. Related problems have been studied under the framework of Combinatorial Multi-Armed Bandits (CMABs), where multiple actions are selected simultaneously from a finite set \cite{chen2016combinatorial,gao2020combinatorial}. CMAB formulations typically assume that the reward distribution of an arm is revealed only upon activation and remains unchanged when the arm is not selected, without explicitly modelling state evolution under passive actions. As discussed in \cite{hao2025combinatorial}, this frozen-arm assumption distinguishes CMABs from restless bandit models. In contrast, the problems considered in this work exhibit the restless property, where each arm evolves according to distinct transition dynamics under both active and passive actions. Consequently, while the action selection is combinatorial, the underlying system dynamics and performance objectives necessitate an \gls{rmab} formulation.

Despite the inherent computational challenges associated with RMABs, the Whittle index has been shown to achieve asymptotically optimal performance for certain classes of problems when the transition probabilities and system dynamics are known \cite{whittle1988restless,xiong2024whittle,wang2023optimistic}. As a result, Whittle-index-based policies have been successfully applied across a wide range of domains, including wireless scheduling, healthcare decision-making, and maintenance optimisation \cite{li2025adaptive,liang2025context,liu2025optimizing}.
Building on this, closed-form Whittle-index-based scheduling policies have been proposed in works such as \cite{kriouile2023pull,chen2021scheduling}, which assume full knowledge of the underlying model parameters. However, such assumptions are often unrealistic in real-world sensing and communication systems, where transition dynamics are unknown or non-stationary \cite{wang2023optimistic,liang2024bayesian}.

To address this limitation, online Whittle index learning approaches for RMABs with unknown dynamics have attracted increasing attention. For example, \cite{avrachenkov2022whittle} proposes a two-timescale Q-learning algorithm that requires careful tuning of multiple learning rates, while \cite{fu2019towards} assumes that the Whittle index lies within a predefined search space and relies on heuristic exploration strategies. Although UCB-based techniques have also been explored \cite{wang2020restless,ortner2012regret}, these approaches typically involve additional hyperparameters and do not directly incorporate a Q-learning formulation.

To the best of our knowledge, this is the first work to propose a fully \gls{ucb}-based Whittle index learning algorithm, \gls{wiql}-\gls{ucb}, that eliminates the need for hyperparameter tuning. We apply this method to the problem of minimising \gls{aoii}, and conduct extensive experiments on benchmark problems to demonstrate its superior performance compared to existing approaches.

In this work, we further develop an online Whittle-index-based scheduling framework for sensor networks by leveraging edge mining techniques to improve state estimation. Rather than transmitting raw sensor data, observations are transformed into application-specific representations at the sensor level, as suggested in \cite{gaura2013edge}. We extend the \gls{ucb}-based approach in \cite{wang2020restless} to support goal-oriented scheduling across diverse application scenarios. Our results demonstrate that near-optimal performance can be achieved without hyperparameter tuning, in contrast to prior methods that rely on carefully selected learning parameters \cite{fu2019towards,avrachenkov2022whittle}. To the best of our knowledge, this is the first work to combine edge mining with online Whittle index learning for goal-oriented, state-aware scheduling in wireless sensor networks.

\section{Preliminaries}
\subsection{Linear dynamic systems }
We consider a sensor monitoring a process governed by a linear dynamic model, denoted as \( x(t) \), defined as:
\begin{equation}
x(t) = A x(t-1) + v(t)
\end{equation}
\noindent
where \( A \) is the update coefficient for the system, and \( v(t) \) represents the system noise.
We denote the measurement received at each time:
\begin{equation}
z(t) = H x(t) + w(t)
\label{eq:state_estimation}
\end{equation}
\noindent
where \( H \) is the observation coefficient that maps the state \( x(t) \) to the measurement \( z(t) \), and \( w(t) \) represents the measurement noise. The state estimate of the linear process can be transformed into \( (x_1, x_2)^T \) as in \cite{gaura2013edge} using the edge mining process:
\begin{equation}
x_1(t) = \beta_1 z(t) + (1 - \beta_1) \left( x_1(t-1) + x_2(t-1) \Delta t \right).
\label{eq:dewma_state}
\end{equation}
\noindent
Similarly, the rate of change \( x_2(t) \) is updated using:
\begin{equation}
x_2(t) = \beta_2 \frac{x_1(t) - x_1(t-1)}{\Delta t} + (1 - \beta_2) x_2(t-1).
\label{eq:dewma_rate}
\end{equation}
where \(\beta_1, \beta_2 \in (0, 1) \) are smoothing factors controlling the influence of the current and previous values,  \( \Delta t \) is the time interval between two consecutive samples. 
The state transformation technique is model-agnostic, allowing the use of various estimation methods such as the \gls{kf} or the normalised least squares technique. The primary requirement is that the same algorithm must be executed on both the node and the sink to maintain consistency in estimation.

\begin{figure*}
    \centering
    \includegraphics[width=\linewidth]{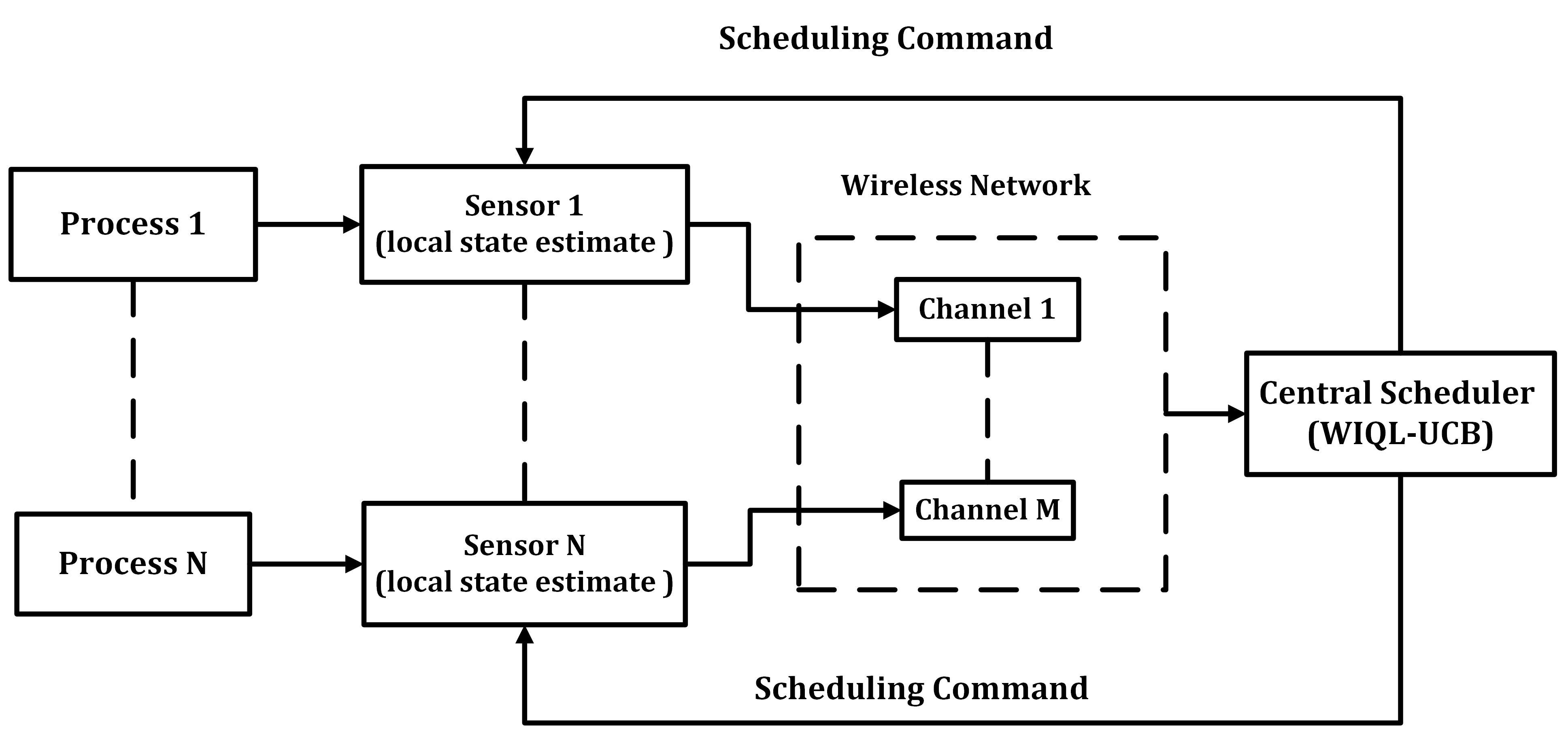}
    \caption{Centralised sensor scheduling framework considered in this work.
Each sensor continuously monitors the underlying physical process and performs local sensing.
Raw measurements are locally transformed into state estimates, but are not transmitted unless the sensor is scheduled.
At each time step, a central scheduler selects \(M\) out of \(N\) sensors subject to channel constraints.
The scheduler runs the \gls{wiql}-\gls{ucb} algorithm, updating its scheduling policy based on the received state updates to optimise the chosen performance objective.}
    \label{fig:centeral_scheduler}
\end{figure*}

\subsection{Sink-Side Estimation of Age of Incorrect Information}

At each time step \( t \), the sink schedules nodes to obtain their state estimates. The polling decision is governed by the action variable \( a_{i,t} \), where \( a_{i,t} \in \{0,1\} \). A value of \( a_{i,t} = 1 \) indicates that the node is polled, while \( a_{i,t} = 0 \) indicates that it is not.
If the transmission is successful, the sink receives a fresh update \( (x_1, x_2)^T \). In the event of a transmission failure or if the node is not polled, the sink estimates the current state based on the last successfully received update at time \( u \), using linear interpolation:
\begin{equation}
\begin{bmatrix}
\hat{x}_1(t) \\
\hat{x}_2(t)
\end{bmatrix}
=
\begin{bmatrix}
1 & t - u \\
0 & 1
\end{bmatrix}
\begin{bmatrix}
x_1(u) \\
x_2(u)
\end{bmatrix}
\label{eq:linear_inter}
\end{equation}
where \( x(u) \) represents the last received state estimate at time \( u \), and \( \hat{x}(t) \) is the predicted state at time \( t \). This prediction follows the assumption of a linear dynamic system, where the state evolves at a constant rate in the absence of new updates. Consequently, the system propagates the state forward in time until a new update is received.

The estimated rate of change \( \hat{x}_2(t) \), multiplied by the time elapsed \( d = t - u \), provides a metric for estimating the deviation of the process since the last update. This metric enables the sink to determine which node, when polled, would provide the most valuable update. 
Unlike the conventional \gls{aoi}, which solely measures the elapsed time \( d \), the distance-based \gls{aoii} formulation captures both the delay and the estimated dynamics of the monitored signal~\cite{kriouile2023pull}. This allows for a more expressive assessment of data staleness, particularly in systems with non-uniform or rapidly changing states.
The change in \gls{aoii} at time \( t \) can be approximated as:
\begin{equation}
    \delta \text{AoII}(t) = d \cdot \hat{x}_2(t)
    \label{eqn:evolving_aoii}
\end{equation}
where \( \hat{x}_2(t) \) denotes the estimated rate of change of the signal at time \( t \).

\subsection{Whittle index for restless bandits}
The \gls{rmab}  is an extension of the classic \gls{mdp}, where in a standard \gls{mdp}, the state evolves solely based on the action taken. In contrast, an \gls{rmab} allows each arm's state to evolve continuously over time even when no action is taken. That is, each arm has distinct transition dynamics under both active and passive actions. Unlike standard Markov problems, where the state of a arm is assumed to be frozen unless acted upon, in an \gls{rmab} the state transitions regardless of whether an action is taken, though the dynamics differ between active and passive modes.

\gls{rmab}s are widely applied in domains requiring efficient allocation of limited resources, such as drug administration in clinical trials, maintenance scheduling, and wireless communication systems. Despite their utility, \gls{rmab} problems are known to suffer from the curse of dimensionality the state space grows exponentially with the number of arms making them PSPACE hard~\cite{papadimitriou1999complexity}. One of the most well known heuristics for addressing this complexity is the Whittle index \cite{whittle1988restless}. The \gls{rmab} tuple is given by the (S,A,R,P)
Given \(N\) arms of which only \(M\) can be activated  and the optimal policy is obtain at each time point the maximum average reward an optimal scheduling policy \( \pi \) that maximises the average total reward across all arms is then formulated as
\begin{equation}
\begin{aligned}
\max_{\pi}\quad & \lim_{T \to \infty} \frac{1}{T} \sum_{t=1}^{T} \sum_{i=1}^{N} \mathbb{E} \left[ R(s_{i,t}, a_{i,t}) \mid \pi \right] \\
\text{subject to}\quad & \sum_{i=1}^{N} a_{i,t} \leq M, \quad \forall t.
\end{aligned}
\label{eq:combined}
\end{equation}
In \cite{whittle1988restless}, Whittle demonstrated that the constrained at every time point:
\begin{equation}
\frac{1}{T} \mathbb{E} \left[ \sum_{t=1}^{T} \sum_{i \in N} a_i^\pi (t) \right] = M.
\label{eq:whittle_constraint}
\end{equation}
By applying Lagrangian relaxation with \( \lambda \) as the Lagrange multiplier and omitting constants, the objective function can be rewritten as:
\begin{equation}
\max_{\pi} \frac{1}{T} \mathbb{E} \left[ \sum_{t=1}^{T} \sum_{i \in N} \left( R(s_{i,t}, a_{i,t}) + \lambda  (1 - a_i^\pi (t)) \right) \right].
\label{eq:whittle_objective}
\end{equation}
With the Whittle formulation, this problem can be decoupled and solved independently for each arm by computing the index \( \lambda_i(s(t)) \). 

\begin{definition}[Indexability]
Let \( \mathcal{D}(\lambda) \subseteq S \) denote the set of states for which it is optimal to take the passive action (i.e., action \( a = 0 \)) when the cost of the active action is \( \lambda \). 

An arm is said to be \textit{indexable} if \( \mathcal{D}(\lambda) \) monotonically increases from the empty set \( \emptyset \) to the full state space \( S \) as \( \lambda \) increases from \( -\infty \) to \( +\infty \). 
An RMAB problem is indexable if all arms in the system are indexable.
\end{definition}

The index is defined such that taking action \( a_i^\pi (t) = 1 \) (activating an arm) is as beneficial as taking action \( a_i^\pi (t) = 0 \) (not activating). 

\begin{equation}
W( s) = \min \left\{ \lambda : Q_{\lambda}(s, 0) = Q_{\lambda}(s, 1) \right\}
\label{eqn:whittle_eqn}
\end{equation}
where \( Q_{c}(s, a) \) and \( V(s) \) are the solutions to the Bellman equation with penalty \( \lambda \) for taking action \( a = 1 \):
\[
Q(s, a) = -\lambda + R(s, a) +  \sum_{s' \in S} P_i(s, a, s') V(s')
\]
\[
V(s) = \max_{a \in A} Q_{c}(s, a)
\]
The optimal penalty, known as the Whittle index \( \lambda(s) \), for each state \( s \in S \), can be computed by solving the following equality:
\begin{equation}
\begin{aligned}
R(s, 1) - \lambda(s) + \sum_{s' \in S} P^1(s, a, s') V(s') \\
= R(s, 0) + \sum_{s' \in S} P^0(s, a, s') V(s').
\end{aligned}
\label{eq:whittle_index_condition}
\end{equation}
where \( P^1 \), \( P^0 \) represents the active and passive transition probability from state \( s \) to \( s' \). Thus from \eqref{eq:whittle_index_condition} the whittle index for each arm in each state is the value of \(\lambda\) such that the active and passive action make no difference in terms of average reward.
This formulation ensures that an arm is only pulled when the expected gain from polling outweighs the associated penalty.
Assuming indexability and known transition probabilities \( P^1 \) and \( P^0 \), the Whittle index \( \lambda_i(s(t)) \) can be computed for each state. Selecting the arms with the highest indices at each time step yields the optimal solution to~\eqref{eq:whittle_objective}.
In most practical cases \( P^1 \) and \( P^0 \) are unknown and techniques such as reinforcement learning can be used to learn the whittle index by observing the state action transition.
By applying the Whittle index technique to a system of \(N\) arms, each with a state space of size \(d\) and two available actions, the task is to learn the Whittle indices for each arm independently. In contrast, standard vanilla Q-learning must learn over the entire joint state-action space of size \((2d)^N\), which becomes intractable for large \(N\).
For example, when \(N = 100\) and each arm has a state space of size \(d = 10\), the Q-learning framework would require learning over:
\[
(2 \cdot 10)^{100} = 20^{100} \text{ state-action pairs}
\]
This exponential growth renders Q-learning computationally infeasible even if neural networks are used as function approximators \cite{avrachenkov2022whittle}.
On the other hand, the Whittle-based approach decomposes the problem into \(N\) independent Q-learning problems. If \(d_1, d_2, \ldots, d_N\) are the cardinalities of the state spaces for the \(N\) arms, then the total number of Q-value updates required under this scheme is:

\[
\sum_{i=1}^{N} \left(2d_i^2 + d_i\right)
\]

Assuming \(d_i = 10\) for all \(i\), the total becomes:

\[
\sum_{i=1}^{100} (2 \cdot 10^2 + 10) = 100 \cdot (200 + 10) = 21{,}000 \text{ updates}
\]
This shows the significant computational advantage of Whittle index-based learning in high-dimensional systems.

\subsection{Q-learning for Whittle index}
Reinforcement learning technique such as Q-learning is popular in learning the optimal policy from state-action interaction without known transition probabilities in  \gls{mdp} simply from it's observation 
\begin{align}
Q(s, a) &= R(s, a) +  \sum_{s' \in S} p^a_{s,s'} 
\max_{a'} Q(s', a'), \label{eq:bellmanQ} \\
V(s) &= \max_{a \in A} Q(s, a) \label{eq:bellmanV}
\end{align}
We assume that the  transition matrices \( P^a \), \( a \in A \), are unknown. The Q-learning algorithm~\cite{watkins1989learning} provides an incremental method to estimate the optimal action-value function. At each time step \( n \), the agent observes the current state \( s \), selects an action \( a \), observes the next state \( s' \), and receives an immediate reward \( R_n = R(s, a) \). The Q-value is then updated as:

\begin{equation}
\begin{aligned}
Q_{n+1}(s, a) &= Q_n(s, a) + \alpha_n \Big( R_n +  
\max_{a'} Q_n(s', a') \\
&\quad - Q_n(s, a) \Big)
\end{aligned}
\end{equation}
for \( (s, a) \in S \times A \); otherwise, \( Q_{n+1}(s, a) = Q_n(s, a) \). Here, \( \alpha_n \) is the learning rate. This is known as \textit{asynchronous} Q-learning, as only one state-action pair is updated at each step. The Q-learning converges to the optimal \( Q^* \) as \( n \to \infty \), provided all state-action pairs are updated infinitely often ~\cite{watkins1989learning,jaakkola1993convergence}.

\subsection{Action Selection Policies}

We discuss two commonly used action selection policies in reinforcement learning: the $\epsilon$-greedy policy and the Upper Confidence Bound (UCB) policy. The convergence speed of Q-learning algorithms is significantly influenced by the choice of the exploration–exploitation strategy.

\paragraph*{$\epsilon$-Greedy Policy}

In the $\epsilon$-greedy policy, the action at each step is selected greedily with probability \( 1 - \epsilon \), based on current Q-value estimates, and selected uniformly at random with probability \( \epsilon \). The policy is formally defined as:

\[
a_n =
\begin{cases}
\text{random action from } A, & \text{with probability } \frac{\epsilon}{|A|}, \\
\arg\max\limits_{a \in A} Q_{n+1}(s, a), & \text{with probability } 1 - \epsilon.
\end{cases}
\]

\paragraph*{UCB-Based Policy}

In the UCB-based action selection strategy, actions are selected to balance exploration and exploitation by adding a confidence bonus to the Q-value. The action at time step \( n \) is chosen as:

\[
a_n = \arg\max_{a \in A} \left[ Q_n(s, a) +   \sqrt{\frac{\log(n + 1)}{N(s, a) + 1}} \right],
\]

where \( N(s, a) \) is the number of times action \( a \) has been taken in state \( s \).

\paragraph*{Q-Value Update}

After selecting action \( a_n \), the next state \( s' \) and reward \( R \) are observed. The Q-value is updated using the standard Q-learning rule:

\[
Q_{n+1}(s, a) = Q_n(s, a) + \alpha_n \left[ R + \gamma \max_{a'} Q_n(s', a') - Q_n(s, a) \right].
\]

\section{Method}
\subsection{Problem Formulation}

Q-Learning  is a reinforcement learning algorithm used to estimate the optimal action-value function \( Q^*(s, a) \) for each state-action pair \( (s, a) \) in a \gls{mdp}. using the Q-learning framework we can directly learn the optimal policy which select the top \(M\) arms to optimise the cummulative average reward at each time point using Algorithm \ref{alg:WIQL-UCB}
\begin{algorithm}[ht]
\caption{Whittle Index Q-Learning with UCB (WIQL-UCB)}
\label{alg:WIQL-UCB}

\KwIn{Number of nodes $N$, polling constraint $M$, learning rate function $\alpha(c)$, initial states $s_i(0) \in S$ for all $i \in [N]$}

\textbf{Initialise:} $Q^0_i(s, a) \gets 0$, $\lambda^0_i(s) \gets 0$, $c^0_{i,s,a} \gets 0$ for all $s$, $a$, and $i$\;

\For{$t = 1, \dots, T$}{
    \tcp{Observe system state and update Q-values}
    \For{$i = 1, \dots, N$}{
        Observe current state $s_i(t)$ and reward $R$\;

        Update visit count:\;
        $c^t_{i,s,a} \gets c^t_{i,s,a} + \mathbb{I}\{s_i(t) = s \text{ and } a_i(t) = a\}$\;

        Compute learning rate: $\alpha(c^t_{i,s,a}) = \frac{1}{1 + c^t_{i,s,a}}$\;

        Update Q-values:\;
        $Q^{t+1}(s, a) \gets (1 - \alpha(c^t_{i,s,a})) Q^t(s, a) + \alpha(c^t_{i,s,a}) \left( R + \max_{a' \in \{0,1\}} Q^t(s', a') \right)$\;

        Compute Whittle index:\;
        $\lambda^{t+1}_i(s) \gets Q^{t+1}_i(s,1) - Q^{t+1}_i(s,0)$\;
    }

    \tcp{Select $M$ arms using UCB-adjusted Whittle indices}
    \For{$i = 1, \dots, N$}{
        Compute UCB term:\;
        $UCB_i(t) \gets \lambda^t_i(s_i(t)) + \sqrt{\frac{2 \log t}{1 + \sum_{a} c^t_{i,s_i(t),a}}}$\;
    }

    Select top $M$ arms with highest $UCB_i(t)$ values and store in $\Psi$\;

    \tcp{Poll only the selected arms}
    \For{$i \in \Psi$}{
        Take active action $a_i(t) = 1$\;
    }
}
\end{algorithm}

\subsection{Theoretical Results}
In this section we want to establish using the optimality of the Q-learning gurantees the optility of the \gls{wiql}-\gls{ucb} policy 
we start by showing that taking the optimal solution that maximises the difference in reward or benefit is equivelent to optimising the joint Q over all arms subject to the constraint \(M\).
\begin{theorem}
\label{thm:one}
Taking action \( a_i = 1 \) on the top \( M \) arms ranked by \( Q^*_i(s_i, 1) - Q^*_i(s_i, 0) \) is equivalent to solving the constrained optimisation problem over all action profiles satisfying \( \sum a_i = M \).
\end{theorem}
Given the optimisation problem under a resource constraint:
\[
\max_{\mathbf{a} \in \{0,1\}^N} \sum_{i=1}^N Q^*_i(s_i, a_i) \quad \text{subject to} \quad \sum_{i=1}^N a_i = M,
\]
where \( Q^*_i(s_i, a_i) \) denotes the expected value of taking action \( a_i \in \{0,1\} \) on arm \( i \) in state \( s_i \), and \( M \) is the maximum number of arms that can be activated at each time step.
To simplify this problem, we define the marginal gain of activating arm \( i \) as:
\[
\Delta_i = Q^*_i(s_i, 1) - Q^*_i(s_i, 0).
\]
Note that for any action \( a_i \in \{0,1\} \), the Q-value can be rewritten as:
\[
Q^*_i(s_i, a_i) = Q^*_i(s_i, 0) + a_i \cdot \Delta_i.
\]
This identity holds because:
\begin{itemize}
    \item If \( a_i = 0 \), then \( Q^*_i(s_i, a_i) = Q^*_i(s_i, 0) \).
    \item If \( a_i = 1 \), then \( Q^*_i(s_i, a_i) = Q^*_i(s_i, 1) = Q^*_i(s_i, 0) + \Delta_i \).
\end{itemize}
Substituting this expression into the original objective, we get:
\begin{align}
\sum_{i=1}^N Q^*_i(s_i, a_i) 
&= \sum_{i=1}^N \left( Q^*_i(s_i, 0) + a_i \cdot \Delta_i \right) \notag \\
&= \sum_{i=1}^N Q^*_i(s_i, 0) + \sum_{i=1}^N a_i \cdot \Delta_i.
\end{align}
The first term \( \sum_{i=1}^N Q^*_i(s_i, 0) \) is constant and independent of the choice of actions. Therefore, maximising the total expected Q-value reduces to maximising the sum of marginal gains:
\[
\max_{\sum a_i = M} \sum_{i=1}^N a_i \cdot \Delta_i.
\]
The solution is obtained by selecting the top \( M \) arms with the highest values of \( \Delta_i \). Thus, a simple greedy policy that selects arms in descending order of \( Q^*_i(s_i, 1) - Q^*_i(s_i, 0) \) is optimal under the constraint.
\begin{theorem}
\label{thm:convergence}
The proposed \gls{wiql}-\gls{ucb} algorithm converges to the optimal solution with probability~1 under the following conditions:
\begin{enumerate}
    \item The learning rate is chosen as
    \[
    \alpha(c)=\frac{1}{1+c},
    \]
    where \( c=c^t_{i,s,a} \) denotes the visit count of state--action pair \( (i,s,a) \).
    \item The exploration policy is UCB-based and each arm induces a communicating Markov decision process.
\end{enumerate}
\end{theorem}

\textit{Proof Sketch.}
The result follows from the classical almost-sure convergence of tabular Q-learning~\cite{watkins1989learning}, which requires: (i) a step-size sequence satisfying the Robbins--Monro conditions, and (ii) infinite visitation of all relevant state--action pairs.

For (i), updates to a given \( (i,s,a) \) occur on its \(c\)-th visit with step size \( \alpha(c)=1/(1+c) \). Since
\[
\sum_{c=1}^{\infty} \alpha(c) = \infty
\quad \text{and} \quad
\sum_{c=1}^{\infty} \alpha(c)^2 < \infty,
\]
the Robbins--Monro conditions are satisfied for each state--action pair. Moreover, sufficient conditions under which visit-dependent (local-clock) learning rates of this form satisfy the Robbins--Monro requirements under persistent exploration have been formally established by Rokhlin~\cite{rokhlin2018robbins}.

For (ii), in the RMAB setting each arm executes an action at every time step: selected arms apply the active action, while non-selected arms apply the passive action. As a result, passive actions are naturally sampled whenever an arm is not scheduled. For active actions, the UCB exploration term assigns higher bonuses to under-sampled state--action pairs, ensuring persistent exploration and preventing any arm--state--action pair from being permanently ignored. Under the assumption that each arm’s underlying Markov dynamics are communicating, this guarantees that all state--action pairs are visited infinitely often with probability one.

Combining (i) and (ii), standard Q-learning convergence results imply that the Q-values \(Q_i^t(s,a)\) converge almost surely to the optimal values \(Q_i^\star(s,a)\) for all arms \(i\), states \(s\), and actions \(a\). Consequently, the learned Whittle indices \(Q_i^t(s,1)-Q_i^t(s,0)\) converge to their optimal counterparts. By Theorem~\ref{thm:one}, selecting the top \(M\) arms according to these indices yields the optimal action profile, and hence \gls{wiql}-\gls{ucb} converges to the optimal policy with probability~1.

\section{Experimental Evaluation}
We compare the performance of our proposed \gls{wiql}-\gls{ucb} algorithm with four benchmark techniques:
\begin{enumerate}
    \item \textbf{Optimal Whittle Index Policy}: In our numerical examples, the Whittle index is computed offline by solving the 
    equation  \eqref{eq:whittle_index_condition}, using the known transition probabilities. While these transition dynamics are not available to the online learning policies, this offline solution serves as an ideal benchmark to evaluate how well the learning algorithms perform relative to the optimal policy.

    \item \textbf{Adaptive $\epsilon$-Greedy Whittle Index Learning}: We implement the approach proposed by Biswas \textit{et al.} \cite{biswas2021learn}, where the \gls{wiql} algorithm adopts both a dynamic $\epsilon$-greedy exploration strategy and an adaptive learning rate \(\alpha_t\) to balance exploration and exploitation while ensuring stable learning. This method is referred to as \gls{wiql}-Biswas in our comparisons.

    \item \textbf{Two-Timescale Whittle Index Q-Learning}: Based on the work of Avrachenkov and Borkar \cite{avrachenkov2022whittle}, we compare against a two-timescale Q-learning approach where the Q-function is updated using a fast learning rate \(\alpha\), while the Whittle index estimate evolves on a slower timescale with learning rate \(B\). This separation allows the Whittle index to stabilise gradually during training. Although the original work assumes homogeneous arms, we consider a more general setting with heterogeneous arms, such as in sensor monitoring tasks. The performance of this method referred to as \gls{wiql}-AB relies  on careful tuning of both \(\alpha\) and \(\beta\), which must be selected specifically for each problem instance to ensure effective learning and convergence.

    \item \textbf{Grid Search-Based Whittle Index Learning}: As proposed by Fu \textit{et al.} \cite{fu2019towards}, this heuristic approach treats the Whittle indices as tunable parameters and performs a grid search over a predefined parameter space to minimise the gap between active and passive value functions. However, this method lacks convergence guarantees if the true optimal Whittle index falls outside the search space, and expanding the search space significantly increases computational cost. We refer to this approach as \gls{wiql}-Fu.

\end{enumerate}
These techniques are evaluated through a series of numerical experiments and practical case studies drawn from the literature, covering diverse application domains to demonstrate the generalisability of the \gls{wiql}-\gls{ucb} approach. 
We further apply \gls{wiql}-\gls{ucb} to a sensor monitoring problem, using both synthetic simulations and a realistic scenario that captures practical sensing and scheduling constraints. The code implementation is available at \underline{https://github.com/sokistar24/whittle\_ucb}.

\subsection{ Circulant dynamics examples}
We begin by evaluating our \gls{wiql}-\gls{ucb} policy on the circulant dynamics benchmark, a well-studied example in the literature for learning the Whittle index using Q-learning-based approaches~\cite{avrachenkov2022whittle, biswas2021learn, fu2019towards}.
In this benchmark, each arm has four states, \( S = \{0, 1, 2, 3\} \), and two possible actions: passive (\( a = 0 \)) and active (\( a = 1 \)). The state transitions are governed by the action selected and follow predefined probability transition matrices:

\begin{figure*}[h]
    \centering
    \begin{subfigure}[t]{0.32\textwidth}
        \centering
         \includegraphics[width=\linewidth]{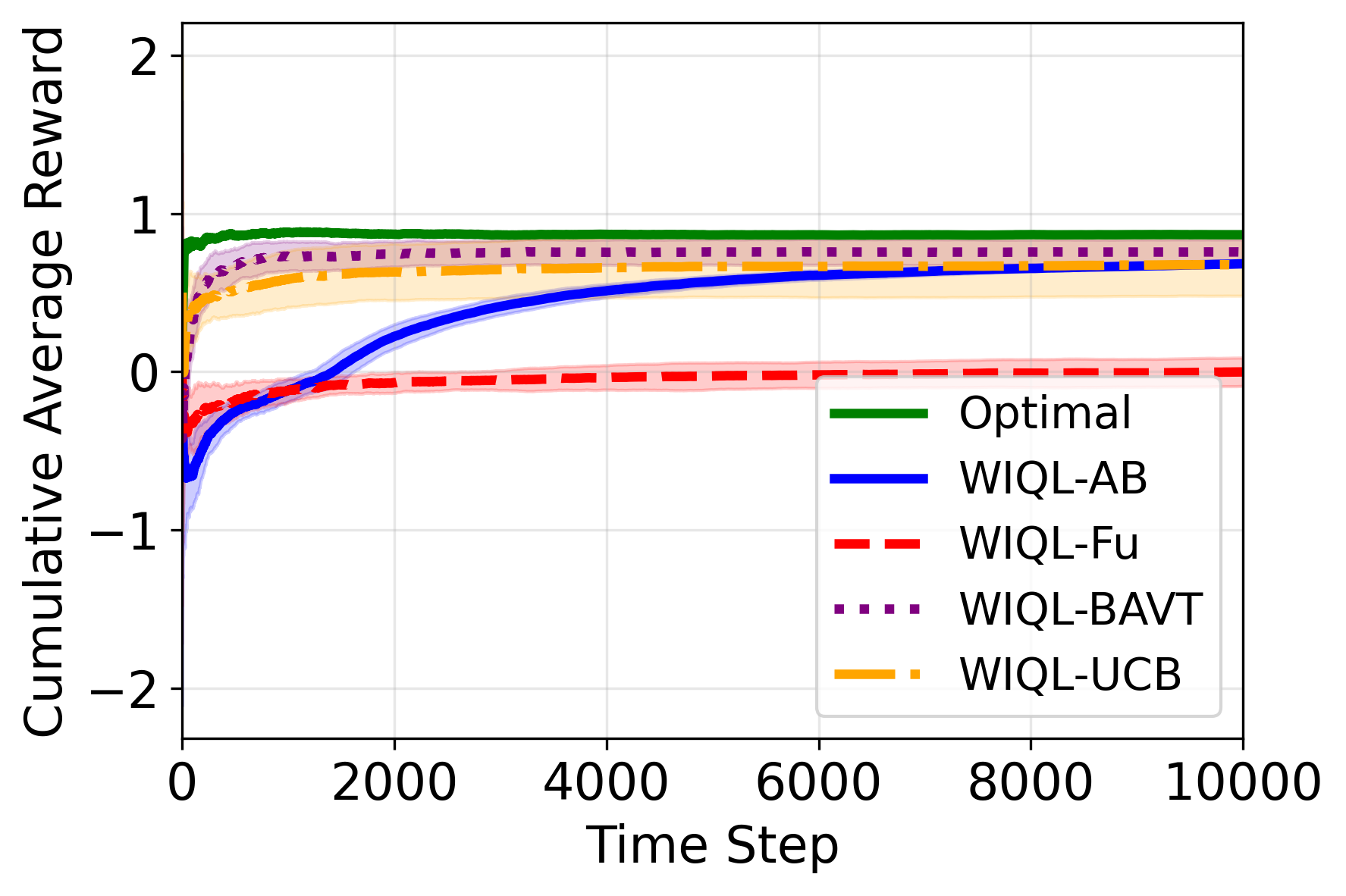}
        \caption{\(N=5\),\( M = 1\)}
        
    \end{subfigure}
    \hfill
    \begin{subfigure}[t]{0.32\textwidth}
        \centering
        \includegraphics[width=\linewidth]{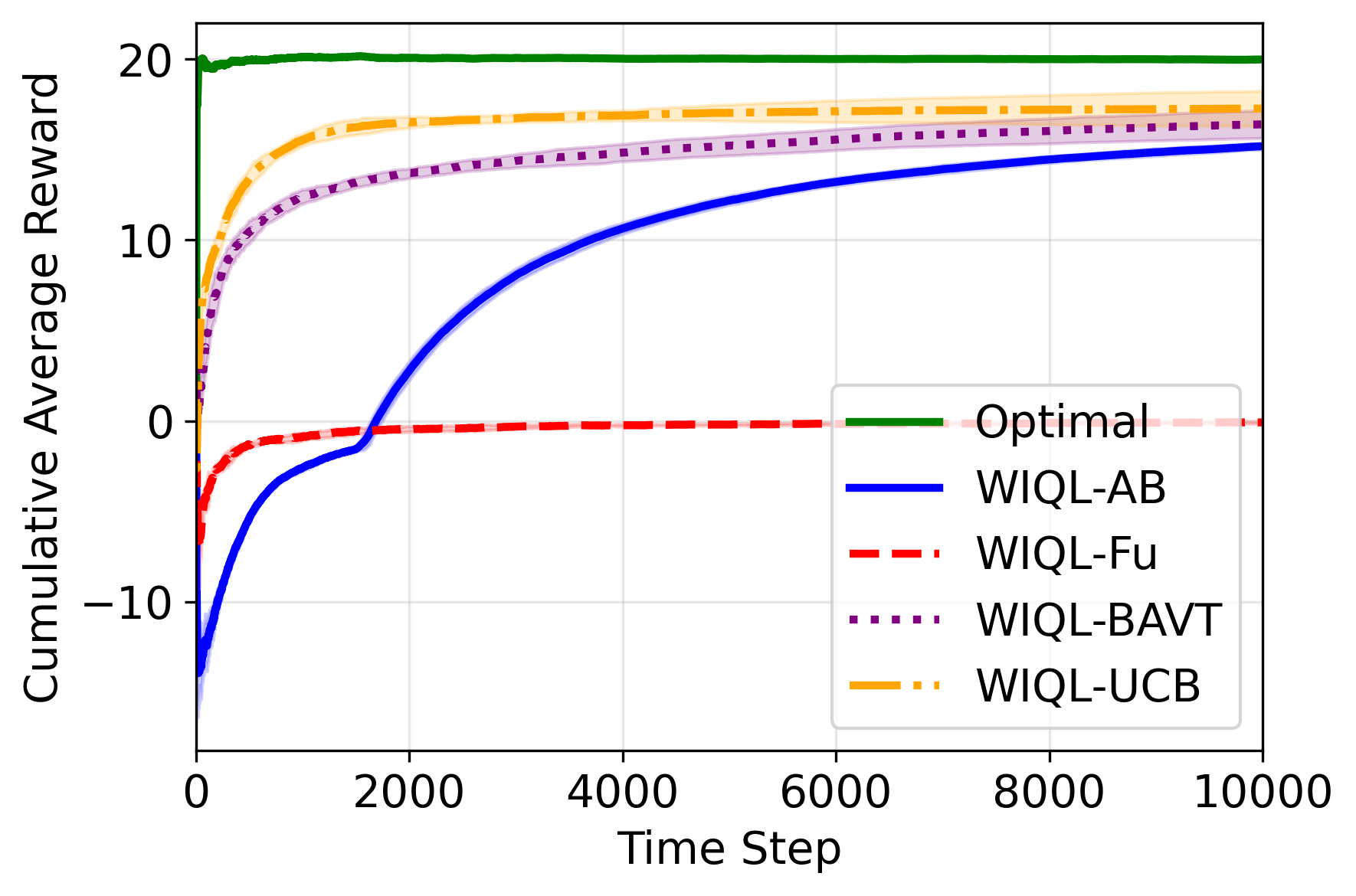}
        \caption{\(N=100\),\( M = 20\)}
        
    \end{subfigure}
    \hfill
    \begin{subfigure}[t]{0.32\textwidth}
        \centering
        
         \includegraphics[width=\linewidth]{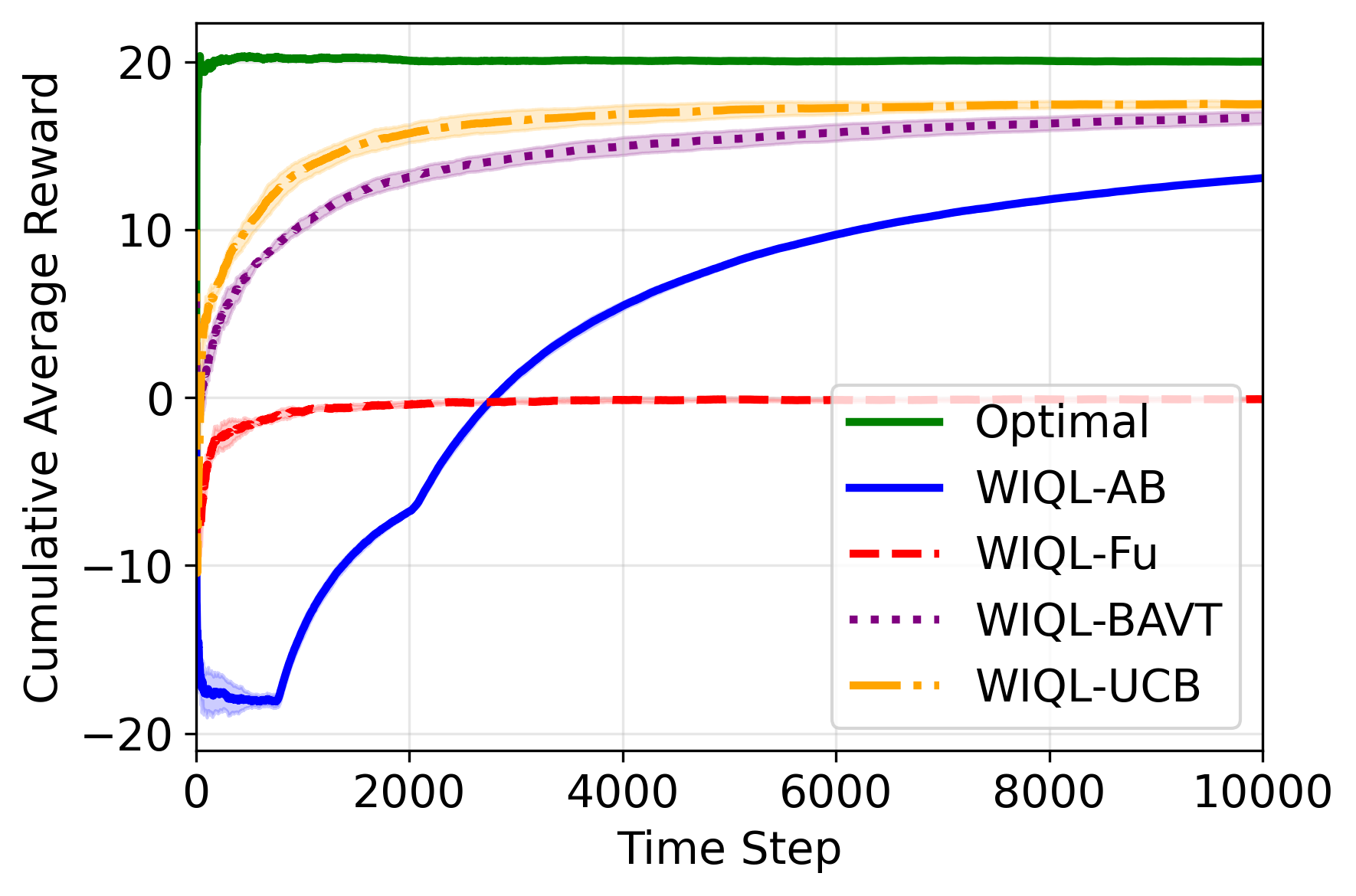}
           \caption{\(N=100\),\( M = 10\)}
        
    \end{subfigure}
\caption{Comparison of average rewards for different scheduling techniques across varying values of \(M\). Each method is evaluated against the oracle optimal policy, which assumes full knowledge of the node-side transition dynamics. In settings with fewer arms, the performance of all policies is relatively similar. However, as the \(N/M\) ratio increases indicating a lower activation budget relative to the number of arms the performance gaps widen. In particular, \gls{wiql}-\gls{ucb} demonstrates superior scalability, achieving performance closest to the oracle benchmark as resource constraints become more pronounced. Experimental results are averaged over 10 simulation runs}

    \label{fig:numerical_example_comp}
\end{figure*}

\[
P_0 =
\begin{bmatrix}
\frac{1}{2} & 0           & 0           & \frac{1}{2} \\
\frac{1}{2} & \frac{1}{2} & 0           & 0           \\
0           & \frac{1}{2} & \frac{1}{2} & 0           \\
0           & 0           & \frac{1}{2} & \frac{1}{2}
\end{bmatrix},
\quad P_1 = P_0^\top.
\]
The rewards depend solely on the current state and are identical for both actions: the reward is \(-1\) in state 0, \(0\) in states 1 and 2, and \(1\) in state 3.
For this numerical example, the optimal Whittle index can be computed offline assuning known transition probabilities using \eqref{eq:whittle_index_condition}
\[
\lambda(0) = -1, \quad \lambda(1) = -\frac{1}{2}, \quad \lambda(2) = \frac{1}{2}, \quad \lambda(3) = 1.
\]
This reward structure intuitively encourages activation when the arm is in state 3, as it provides the highest possible reward.

Fig.~\ref{fig:numerical_example_comp} presents the performance comparison. In scenarios with a small number of arms (\(N = 5\)) and a low activation ratio (\(M = 1\)), all algorithms perform comparably and closely approximate the optimal policy, with the exception of \gls{wiql}-Fu. This is consistent with the findings in \cite{fu2019towards}, where \gls{wiql}-Fu converges to an average reward of approximately 0.08.
However, as the problem scales to larger systems with \(N = 100\) and reduced activation budgets (\(M = 20\) and \(M = 10\)), the performance differences become more pronounced. Specifically, as the budget constaints gets tighter either \(M\) reduces relative to \(N\) , the performance of \gls{wiql}-AB deteriorates and its convergence slows. It is worth noting that the experimental settings for Fig.~\ref{fig:numerical_example_comp}a and Fig.~\ref{fig:numerical_example_comp}b are aligned with those used in \cite{avrachenkov2022whittle, biswas2021learn}.
In these larger-scale settings, \gls{wiql}-AB deviates further from the optimal policy, while \gls{wiql}-\gls{ucb} consistently achieves the best performance. Notably, the \gls{ucb}-based Whittle Index Q-Learning method outperforms other approaches as the ratio \(N/M\) increases, highlighting its scalability and robustness in resource-constrained environments.

\subsection{Process update example }
\label{section:process_update}

In this example, we consider a setting where the active action causes an arm to restart from a specific state. This behaviour is typical in \gls{aoii}-based semantic applications, where a successful update from a sensor node resets the \gls{aoii} at the remote monitor \cite{ayik2023optimization,kriouile2021minimizing,kriouile2023pull}. It is also relevant in other domains such as congestion control and machine maintenance, as discussed in \cite{avrachenkov2022whittle}.
A similar benchmark has been studied in the literature, notably in \cite{avrachenkov2022whittle}, and is also reproduced in Appendix~(\ref{appendix:restart_example}). However, in this work, we extend the setting by assuming the arms are heterogeneous, i.e., the some of the arms have different transition probabilities.
Furthermore, we explore two distinct scenarios: a static case, where the transition probabilities remain constant, and a dynamic case, where the transition probabilities of the underlying Markov processes change during the simulation.
In this setting, each arm has five states, indexed from 0 to 4, representing the freshness of information state 0 being the most updated and state 4 the most outdated. When an arm is in the passive state, it may remain in the same state with some probability, reflecting situations where the state of the underlying system does not change. The transition probabilities under the passive action differ across three categories of arms: A, B, and C.

When an arm is activated (i.e., the active action is taken), it attempts to transmit an update. As in sensing application, this transmission may succeed or fail as in wireless sensors due to imperfect transmission channel. If successful, the \gls{aoii} resets to state 0; if not, the state may remain unchanged or degrade further. Each arm category has different probabilities of remaining in the same state (\(P_s\)) versus transitioning to a new state (\(P_r\)) upon activation:
Category A: \(P_s = 0.6\), \(P_r = 0.4\)
Category B: \(P_s = 0.9\), \(P_r = 0.1\)
Category C: \(P_s = 0.5\), \(P_r = 0.5\)

\vspace{1em}
\begin{figure*}[!h]
    \centering
    \begin{subfigure}[t]{0.32\textwidth}
        \centering
        \includegraphics[width=\linewidth]{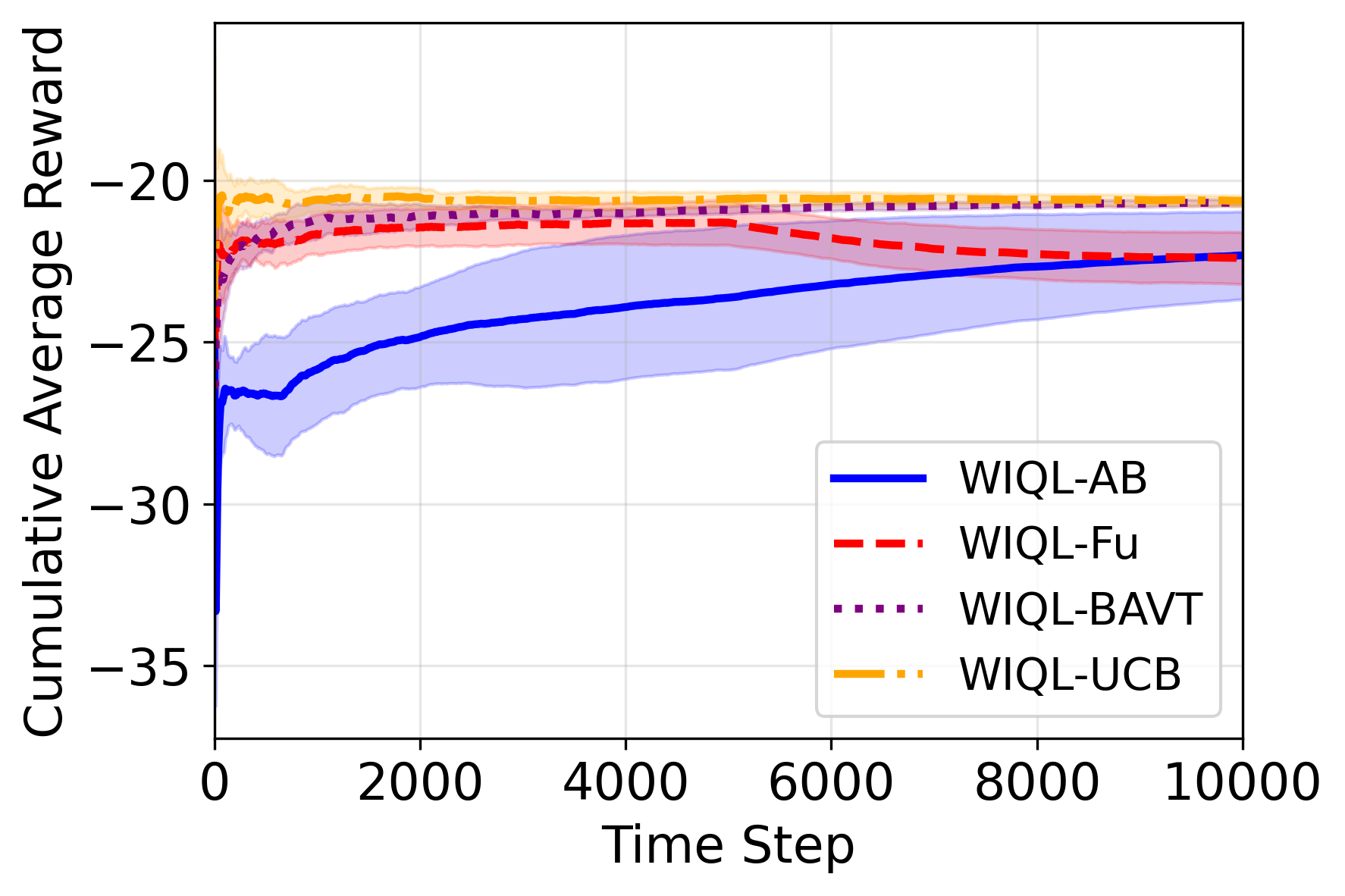}
        \caption{Static: \( N = 12 \), \( M = 1 \)}
    \end{subfigure}
    \hfill
    \begin{subfigure}[t]{0.32\textwidth}
        \centering
        \includegraphics[width=\linewidth]{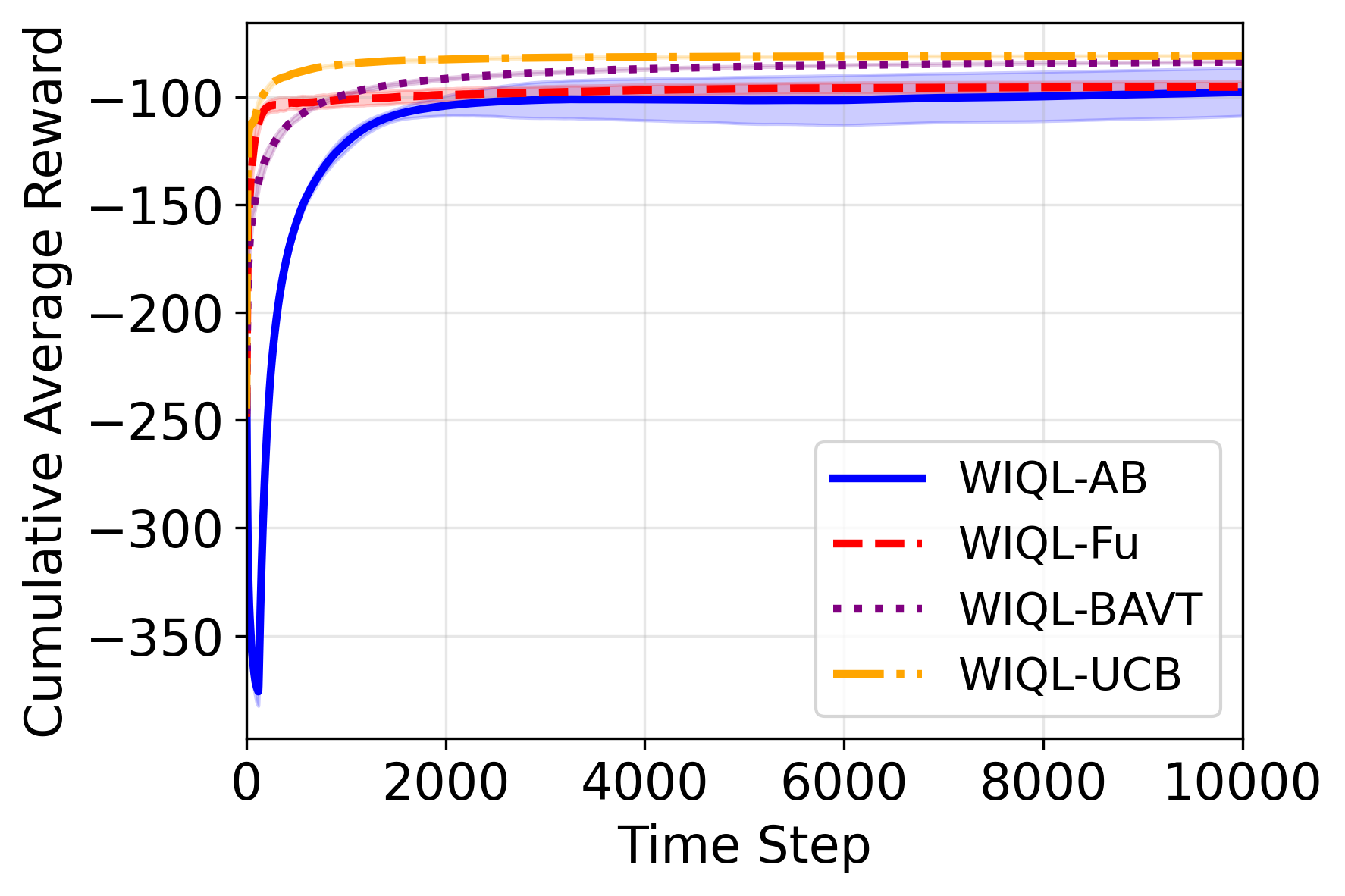}
        \caption{Static: \( N = 120 \), \( M = 20 \)}
    \end{subfigure}
    \hfill
        \begin{subfigure}[t]{0.32\textwidth}
        \centering
        \includegraphics[width=\linewidth]{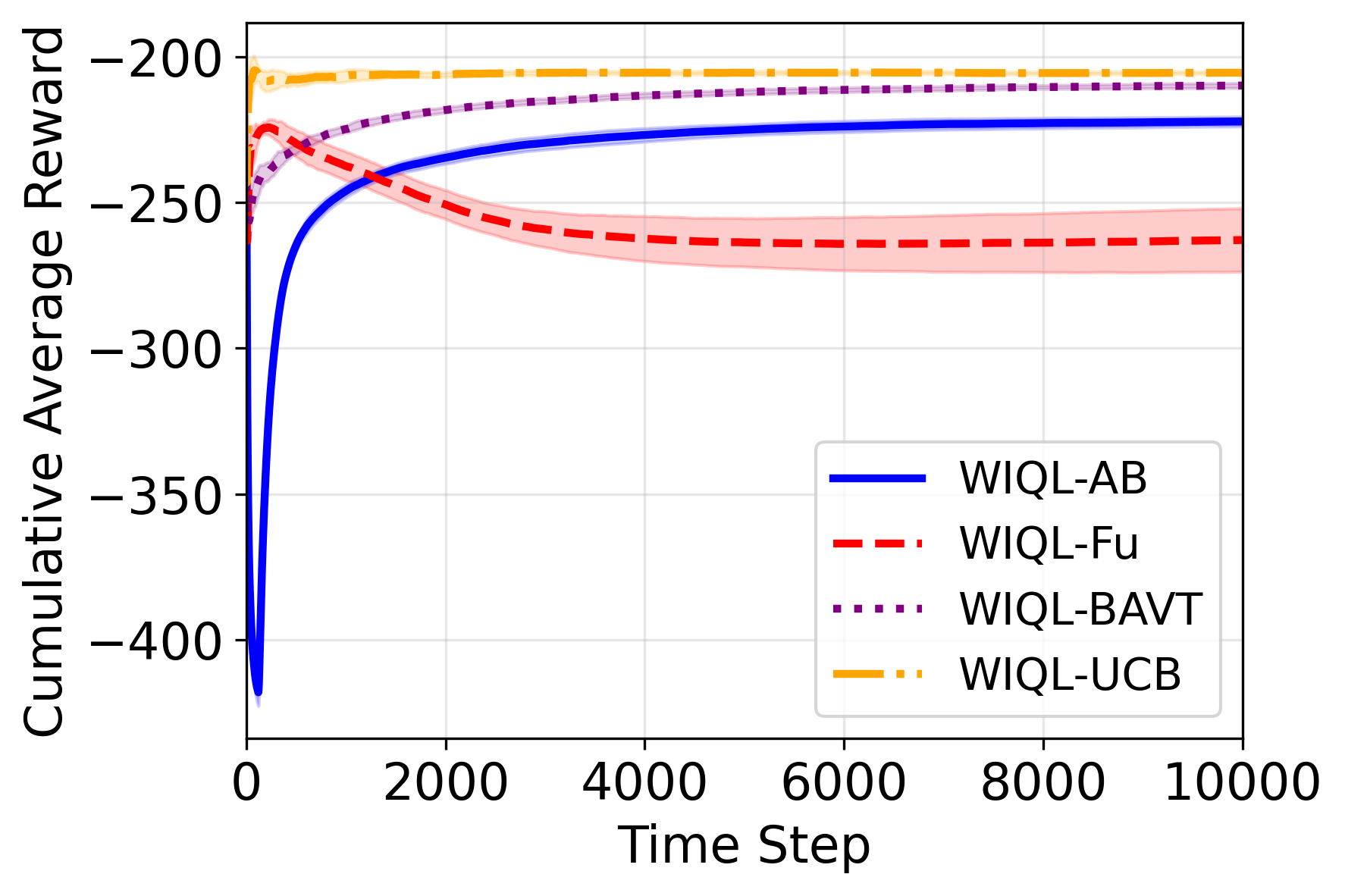}
        \caption{Static: \( N = 120 \), \( M = 10 \)}
    \end{subfigure}
    \vspace{1em} 

    \begin{subfigure}[t]{0.32\textwidth}
        \centering
        \includegraphics[width=\linewidth]{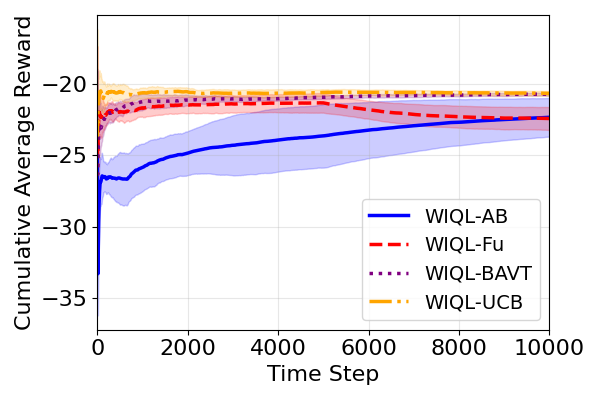}
        \caption{Dynamic: \( N = 12 \), \( M = 1 \)}
    \end{subfigure}
    \hfill
       \begin{subfigure}[t]{0.32\textwidth}
        \centering
        \includegraphics[width=\linewidth]{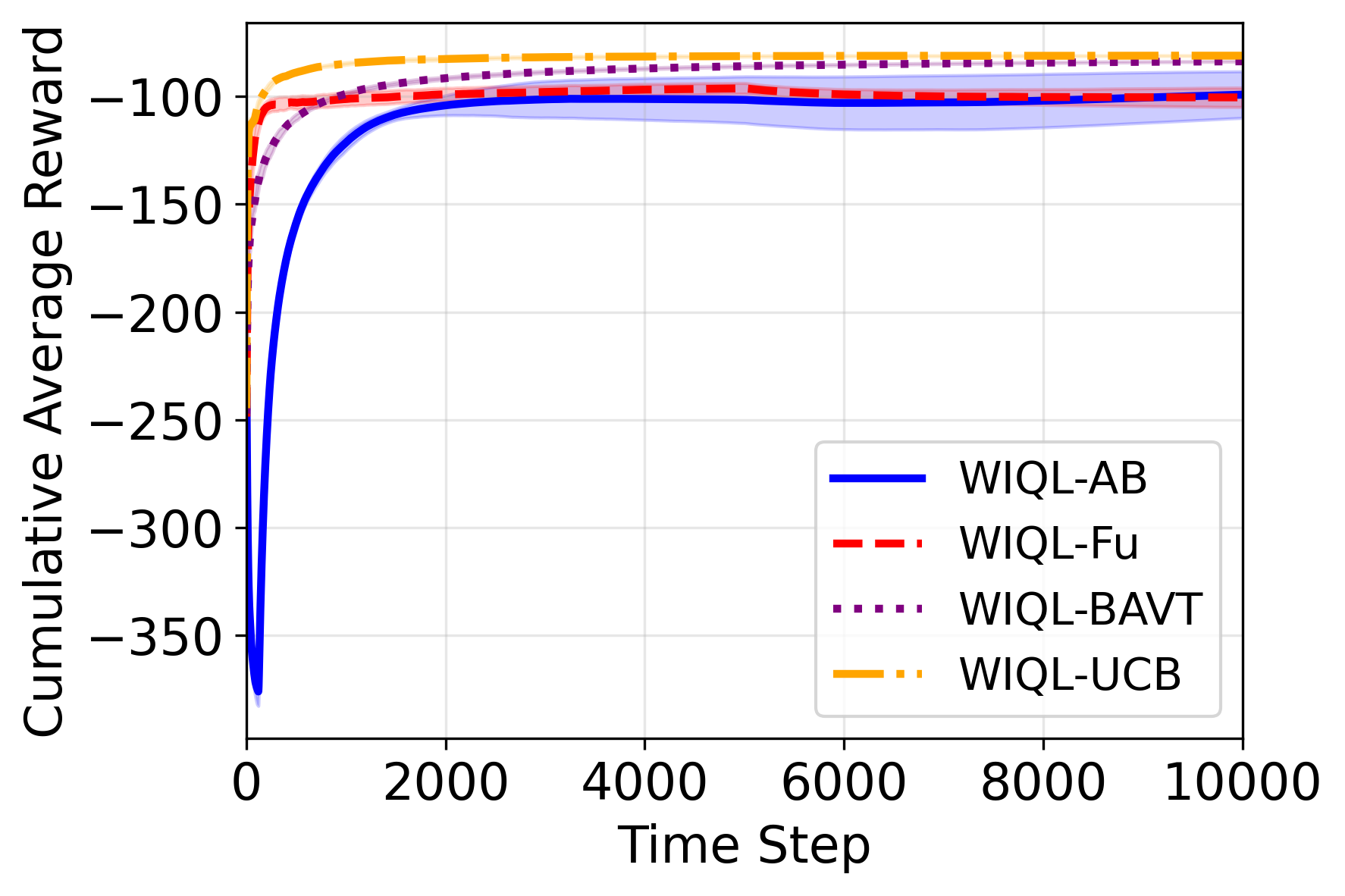}
        \caption{Dynamic: \( N = 120 \), \( M = 20 \)}
    \end{subfigure}
    \hfill
 \begin{subfigure}[t]{0.32\textwidth}
        \centering
        \includegraphics[width=\linewidth]{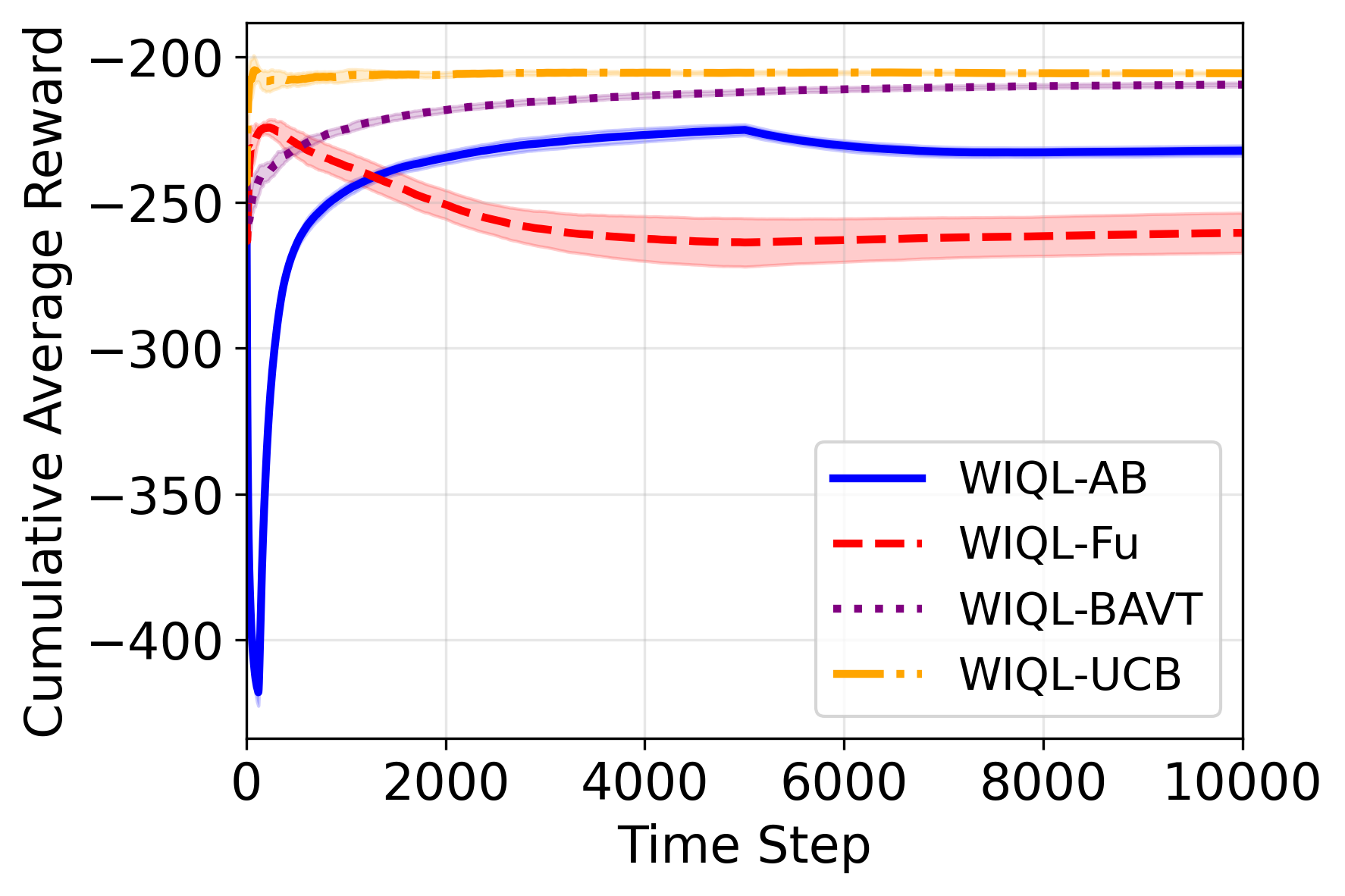}
        \caption{Dynamic: \( N = 120 \), \( M = 10 \)}
    \end{subfigure}
    \caption{Comparison of average rewards for different scheduling techniques in static (top row) and dynamic (bottom row) environments, across varying values of \(M\) and \(N\). \gls{wiql}-\gls{ucb} consistently outperforms the other strategies, particularly as system constraints increase. In contrast, \gls{wiql}-Fu and \gls{wiql}-AB struggle to adapt in dynamic environments due to their reliance on outdated transition probability estimates and slower adaptation to changing system dynamics.}

    \label{fig:combined_static_dynamic}
\end{figure*}

\textit{Passive action transition matrix:}
\[
P^{\text{passive}}_A =
\begin{bmatrix}
0.6 & 0.4 & 0.0 & 0.0 & 0.0 \\
0.0 & 0.6 & 0.4 & 0.0 & 0.0 \\
0.0 & 0.0 & 0.6 & 0.4 & 0.0 \\
0.0 & 0.0 & 0.0 & 0.6 & 0.4 \\
0.0 & 0.0 & 0.0 & 0.0 & 1.0 \\
\end{bmatrix}
\]

\textit{Active action transition matrix:} When an arm is activated, there is a 90\% chance it resets to state 0 (successful transmission), and a 10\% chance it remains in the same state (packet loss), except in state 0, where activation has no effect.

\[
P^{\text{active}}_A =
\begin{bmatrix}
1.0 & 0.0 & 0.0 & 0.0 & 0.0 \\
0.9 & 0.1 & 0.0 & 0.0 & 0.0 \\
0.9 & 0.0 & 0.1 & 0.0 & 0.0 \\
0.9 & 0.0 & 0.0 & 0.1 & 0.0 \\
0.9 & 0.0 & 0.0 & 0.0 & 0.1 \\
\end{bmatrix}
\]
Similar transition matrices are defined for categories B and C, with different probabilities to reflect varying update dynamics and reliability across arms.
For the reward structure, we assume that outdated information is penalised, with lower rewards assigned to higher AoII states. Specifically, the reward is defined as the negative of the state index:
\[
R(s) = -s.
\]
In the \textit{static} scenario, the transition probabilities for all arms remain constant throughout the entire simulation run. In contrast, the \textit{dynamic} scenario models a non-stationary environment: at the halfway point of the simulation, the system dynamics of categories A and B are interchanged. While this change is deterministic in our setup, it reflects the fact that in real-world systems, such changes often occur arbitrarily. This scenario highlights the need for scheduling policies that can adapt to evolving system dynamics.

As illustrated in Fig.~\ref{fig:combined_static_dynamic}, when the ratio \(N/M\) is low either because \(M\) is relatively large (i.e., more arms can be activated) or the scheduling constraint is less severe the performance gap between the methods remains marginal across both static and dynamic settings. However, as the \(N/M\) ratio increases and the system becomes more constrained (i.e., fewer arms can be activated), the performance differences become more pronounced. In particular, the performance of \gls{wiql}-Fu and \gls{wiql}-AB degrades due to their limited adaptability to the changing transition dynamics, whereas the \gls{wiql}-\gls{ucb} policy consistently outperforms all other approaches.

It is worth noting that the performance of \gls{wiql}-Fu and \gls{wiql}-AB degrades significantly under tighter constraints (e.g., low \(M\) with fixed \(N\)). This is primarily due to their use of a shared Q-table across all arms, which limits their ability to capture arm-specific dynamics. While such a design may be suitable when the arms are homogeneous, it becomes a major limitation in heterogeneous or dynamic environments. As a result, these methods struggle to adapt their learning rates and exploration-exploitation strategies to the varying behaviours of individual arms, leading to suboptimal performance in more complex settings.

\subsection{Environmental Monitoring Application}
We further extend our evaluation by comparing the proposed \gls{wiql}-\gls{ucb} policy against commonly used baseline techniques in sensor monitoring, particularly in constrained network settings. Specifically, we compare with standard \gls{rr} and \gls{aoi}-based scheduling\cite{wang2021sleep,long2024aoi,zhu2021aoi}.
The \gls{rr} policy selects each node (or arm) in a fixed cyclic order, pulling them periodically regardless of their information freshness. In contrast, the \gls{aoi}-based policy selects the node with the largest delay since its last successful transmission, denoted by \( d \), aiming to reduce the staleness of information.
This problem can be formulated as a \gls{rmab}, where only \( M \) out of \( N \) nodes can be selected (or polled) at each time step to update their state. 
Recall from \eqref{eqn:evolving_aoii} the evolution of the change in \gls{aoii} is given by:
\begin{equation}
    \delta \gls{aoii}(t) =  d \cdot \hat{x}_2(t),
    \label{eqn:evolving_aoii_recall}
\end{equation}
where \( \hat{x}(t) \) represents the estimated error at time \( t \) from the remote monitor or sink.
This setting is analogous to the process update scenario described in Section~\ref{section:process_update}, where increasing \gls{aoii} corresponds to a less favourable system state (i.e., a higher penalty or lower reward). Hence, a pulling action becomes desirable to reset the AoII and improve the system's information accuracy.
Although the AoII in this formulation is continuous, assuming known upper bounds for both \( d \) and \( \hat{x}(t) \), the continuous state space can be discretised. This enables the application of the proposed \gls{wiql}-\gls{ucb} framework for learning near-optimal scheduling policies in this setting.

For this example, we consider a temperature monitoring application involving three class of sensors monitoring and environment which could represent typical industrial machinery or remote environmental monitoring were some of the sensors are measuring various environments. These sensors exhibit different rates of change, which we classify as category  \(A\) slow, category  \(B\) medium, and category \( C\) fast.
The temperature dynamics for each sensor $i$ at time step $t$ are given by
\begin{equation}
    T_i(t) = 20 + A_i \sin\left(\frac{2\pi t}{P_i}\right) + \mathcal{N}(0, \sigma_i)
\end{equation}

\noindent where:
\begin{itemize}
    \item $T_i(t)$ is the temperature reading of sensor $i$ at time $t$,
    \item $A_i$ is the amplitude of the temperature variation,
    \item $P_i$ is the period of oscillation for sensor $i$,
    \item $\mathcal{N}(0, \sigma_i)$ represents the Gaussian noise with zero mean and standard deviation $\sigma_i$.
\end{itemize}
We assume a scenario were different sensors monitors different with varying rate of change in the environment:
\begin{itemize}
    \item \textbf{Category A:} $i \in \{1,2, \dots, 10\}$ 
    \begin{equation}
        P_i = 500, \quad \sigma_i = 0.2
    \end{equation}
 
    \item \textbf{Category B:} $i \in \{11,12, \dots, 20\}$ 
    \begin{equation}
        P_i = 200, \quad \sigma_i = 0.3
    \end{equation}
       \item \textbf{Category C:} $i \in \{21,22, \dots, 30\}$ 
    \begin{equation}
        P_i = 50, \quad \sigma_i = 0.5
    \end{equation}
\end{itemize}

\begin{figure*}[!h]
    \centering
    \begin{subfigure}[t]{0.32\textwidth}
        \centering
        \includegraphics[width=\linewidth]{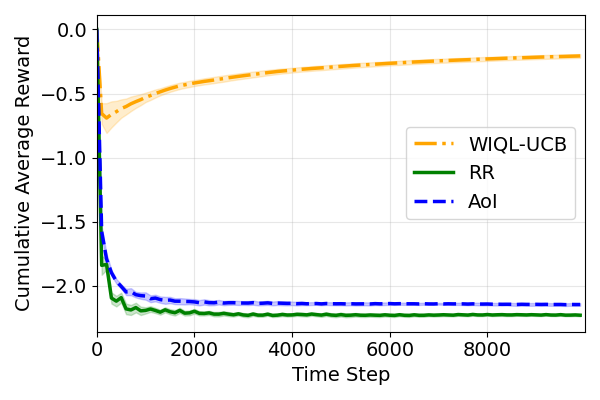}
        \caption{\( M = 1\)}
        
    \end{subfigure}
    \hfill
    \begin{subfigure}[t]{0.32\textwidth}
        \centering
        \includegraphics[width=\linewidth]{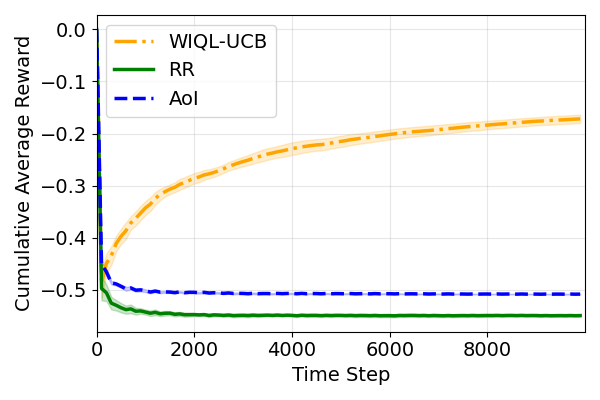}
        \caption{\(M = 5\)}
        
    \end{subfigure}
    \hfill
    \begin{subfigure}[t]{0.32\textwidth}
        \centering
        
         \includegraphics[width=\linewidth]{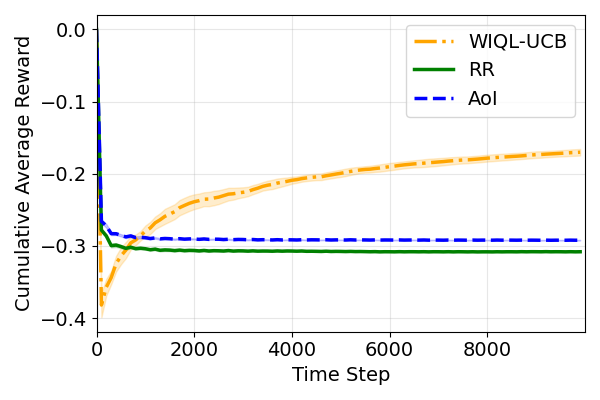}
         \caption{\(M = 10\)}
        
    \end{subfigure}
  \caption{Comparison of rewards for various scheduling techniques under different values of \(M\). \gls{wiql}-\gls{ucb} consistently achieves the highest reward across all settings. The performance gap is most pronounced when the ratio \(N/M\) is high, and gradually decreases as the number of activations \(M\) increases.}

    \label{fig:varying_M_constraint}
\end{figure*}

\begin{figure}[h]
    \centering
    \includegraphics[width=\linewidth]{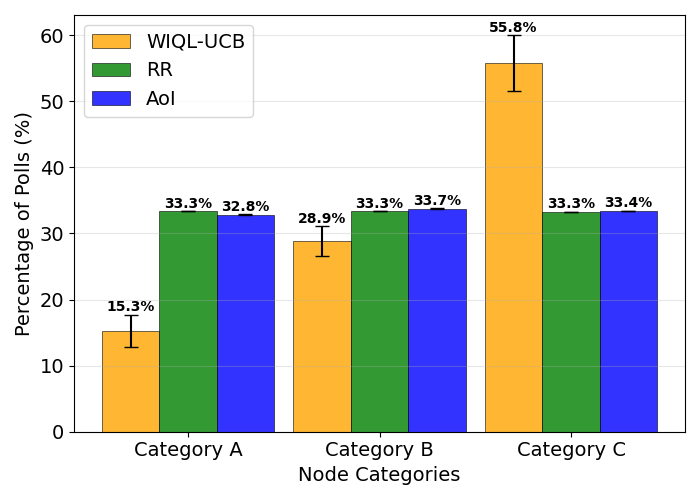}
  \caption{Node pulling comparison across categories. \gls{wiql}-\gls{ucb} activates nodes in Category C most frequently, aligning activation priority with variation level. In contrast, \gls{rr} and \gls{aoi} allocate pulls uniformly across categories, ignoring differences in variation.}

    \label{fig:poll_distribution}
\end{figure}

We compare the performance of our proposed \gls{wiql} policy with baseline techniques such as \gls{rr} and \gls{aoi}-based polling. Ideally, to minimise \gls{aoii} and ensure the sink remains updated with the most recent sensor readings, the scheduling policy should prioritise pulling sensors whose system states evolve more rapidly ensuring that the remote monitor maintains the most accurate view of the environment. This approach differs fundamentally from \gls{aoi}-based methods, which update solely based on time elapsed, without accounting for the actual state evolution of the monitored system.

To capture this behaviour, we define the reward as the negative of the \gls{aoii}: lower \gls{aoii} corresponds to a higher reward, while higher \gls{aoii} results in a lower reward. As shown in Fig.~\ref{fig:varying_M_constraint}, \gls{wiql}-\gls{ucb} consistently achieves the highest average reward across varying scheduling constraints \(M\), outperforming the baseline methods.
As illustrated in Fig.~\ref{fig:poll_distribution}, the \gls{wiql}-\gls{ucb} policy allocates the highest polling frequency to nodes in Category C those with the fastest changing state dynamics demonstrating its ability to prioritise nodes that require more frequent updates. This behaviour reflects the core principle of an effective scheduling policy: prioritising rapidly evolving systems to maintain accurate and timely state information at the sink.
In contrast, baseline methods such as \gls{rr} and \gls{aoi} assign equal sampling opportunities to all nodes based solely on elapsed time, regardless of the underlying system dynamics. This leads to inefficiencies, as slow-changing nodes are polled as often as fast-changing ones, resulting in less informative updates and suboptimal use of limited communication resources.
We next consider sensing data from a real-world wireless sensor deployment using the publicly available Intel Berkeley Research Laboratory dataset~\cite{madden2010}. This dataset comprises approximately 2.3 million measurements collected from 54 sensor nodes deployed throughout a large indoor environment. The sensors record multiple physical quantities, including temperature, humidity, voltage, and light intensity. Data were collected continuously over a four-month period using Mica2Dot sensor motes and the TinyDB in-network query processing system~\cite{madden2010}.

\section{Comparison of Non Whittle Index Approach}

This section compares the performance of non–Whittle-index scheduling policies with the proposed \gls{wiql}-\gls{ucb} approach. Specifically, a greedy policy, vanilla Q-learning, \gls{dqn} and \gls{ppo} are evaluated against the standard \gls{wiql}-\gls{ucb} policy across the circulant dynamics problem and a restart problem based on the \gls{rmab} framework of Avrachenkov and Borkar~\cite{avrachenkov2022whittle}, in which selecting the active action forces an arm to reset to its initial state.

In small-scale settings with limited state–action spaces, such as \(N = 5, M = 1\), the non-Whittle-index methods achieve competitive performance. In particular, the \gls{dqn} policy performs close to the optimal solution and exhibits behaviour comparable to that of the \gls{wiql}-\gls{ucb} policy.
However, as \(N\) and \(M\) increase thereby substantially enlarging the state–action space the performance of non–Whittle-index methods degrades. In larger-scale settings such as \(N = 10, M = 3\) and \(N = 20, M = 4\), their performance approaches that of the greedy policy, indicating limited scalability. This behaviour is consistent with theoretical expectations, which show that the complexity of the problem grows exponentially with the size of the state–action space

\begin{figure*}[h]
  
  \subfloat[\(N=5,\, M=1\)]{%
    \includegraphics[width=0.32\linewidth]{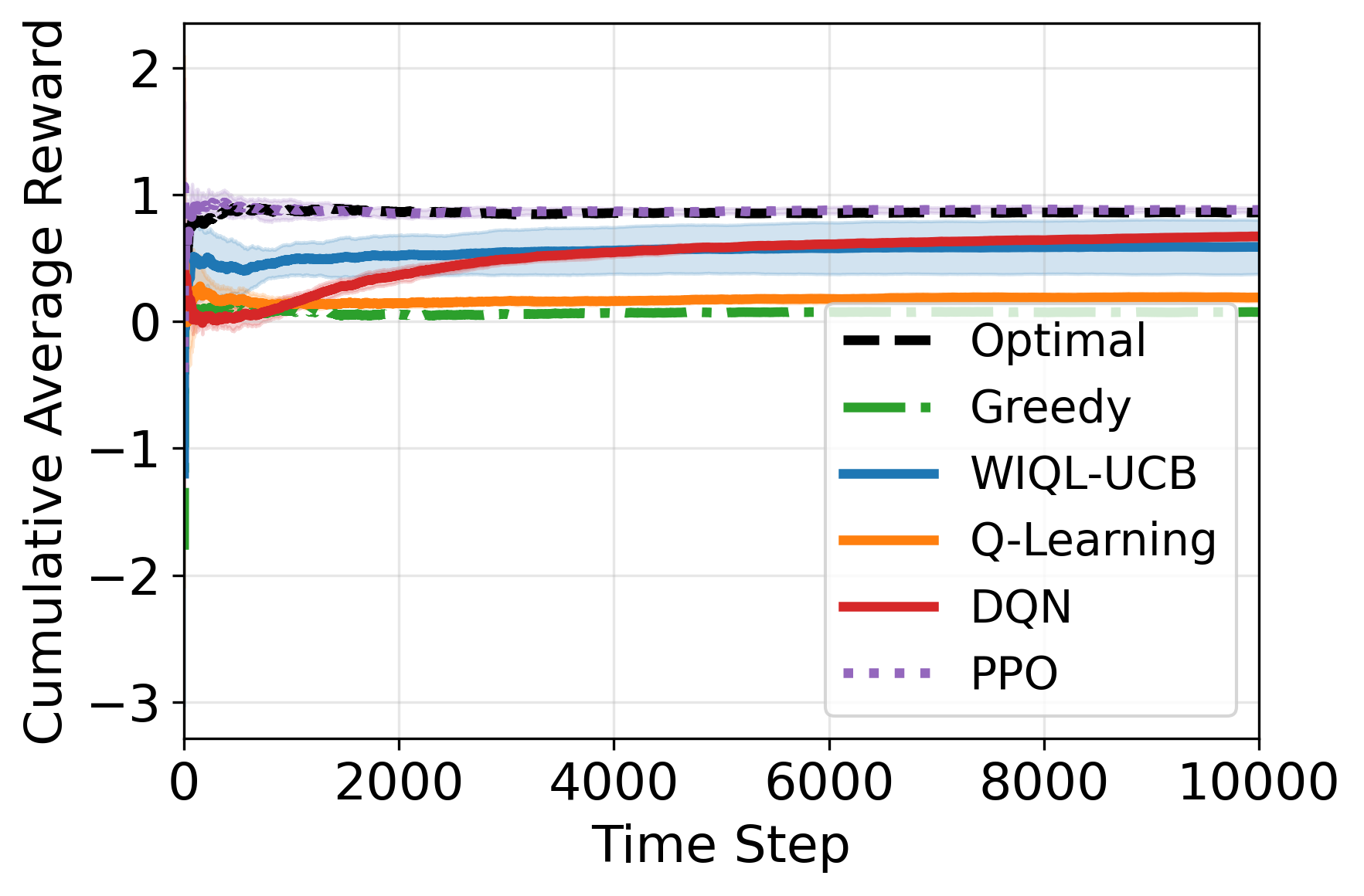}
   
  }
  \subfloat[\(N=15,\, M=3\)]{%
    \includegraphics[width=0.32\linewidth]{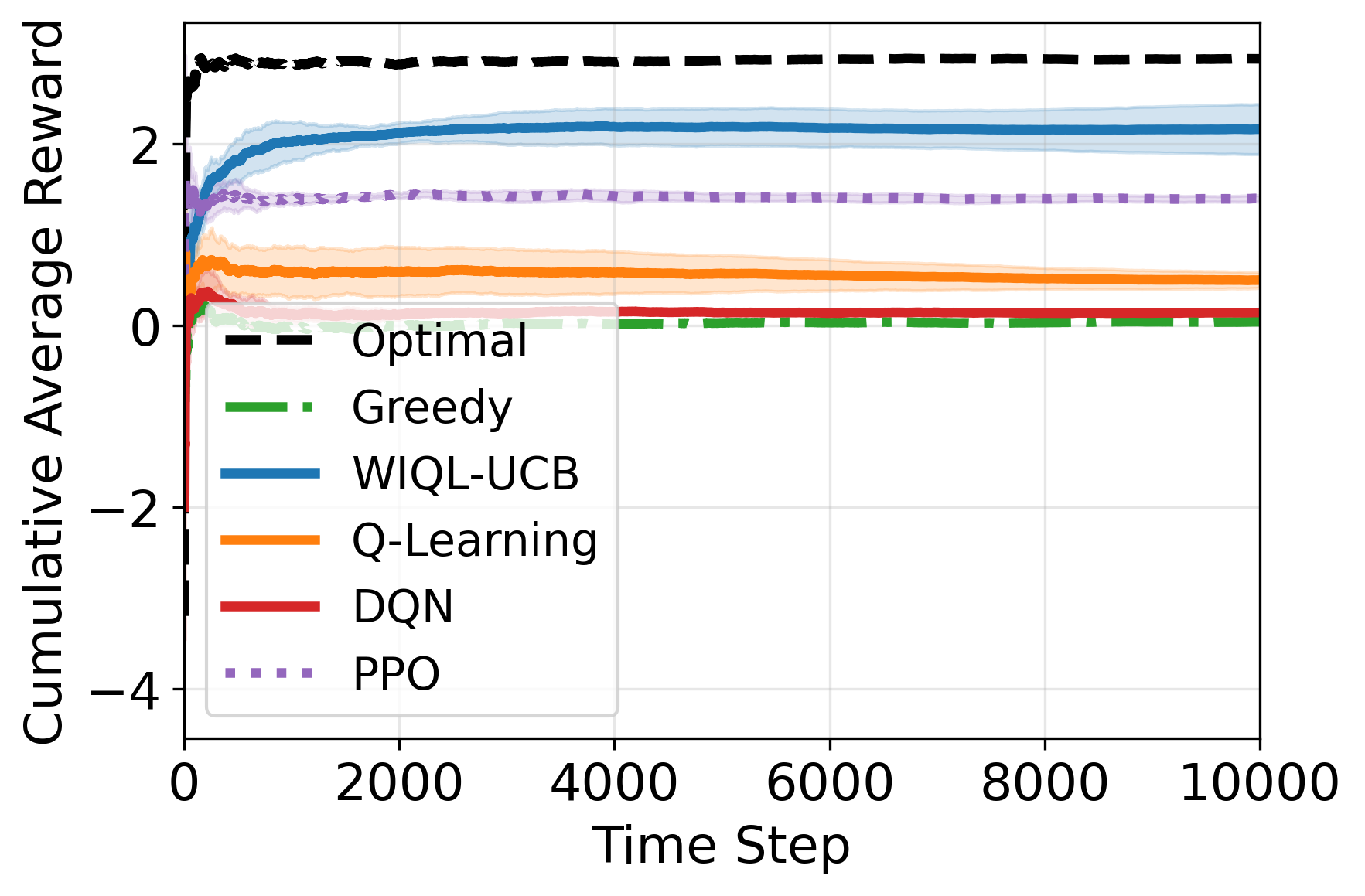}
    
  }
  \subfloat[\(N=20,\, M=4\)]{%
    \includegraphics[width=0.32\linewidth]{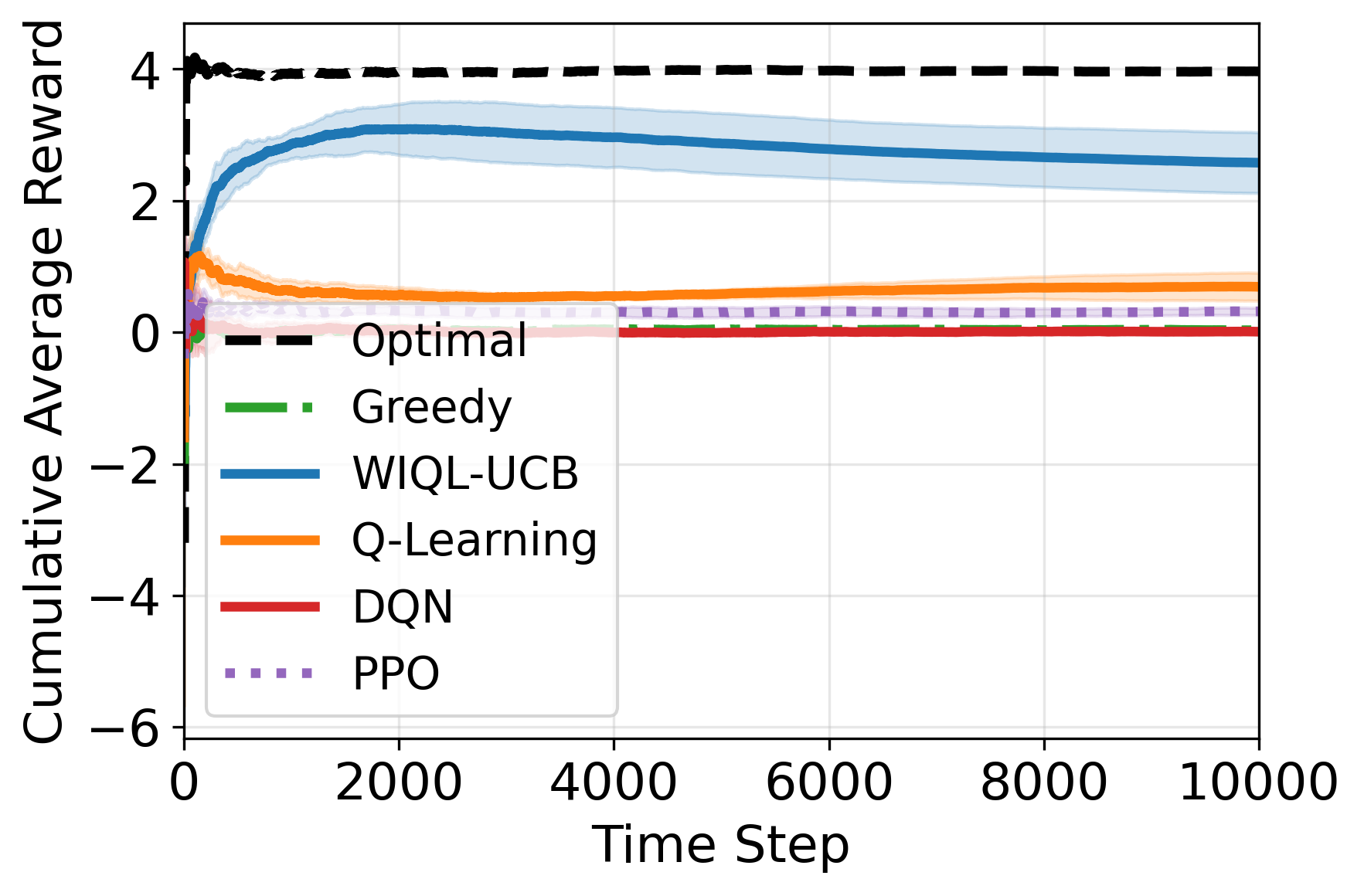}
    
  }

 \caption{Circulant Dynamics: Comparison of average rewards for Whittle-index and non–Whittle index scheduling policies across increasing problem sizes. While non–Whittle index methods perform competitively for small state action spaces, their performance degrades as \(N\) and \(M\) increase; \gls{wiql}-\gls{ucb} remains closest to the oracle.}
  \label{fig:Whittle_and_non_index_baselines}
\end{figure*}

\begin{figure*}[h]
  
  \subfloat[\(N=5,\, M=1\)]{%
    \includegraphics[width=0.32\linewidth]{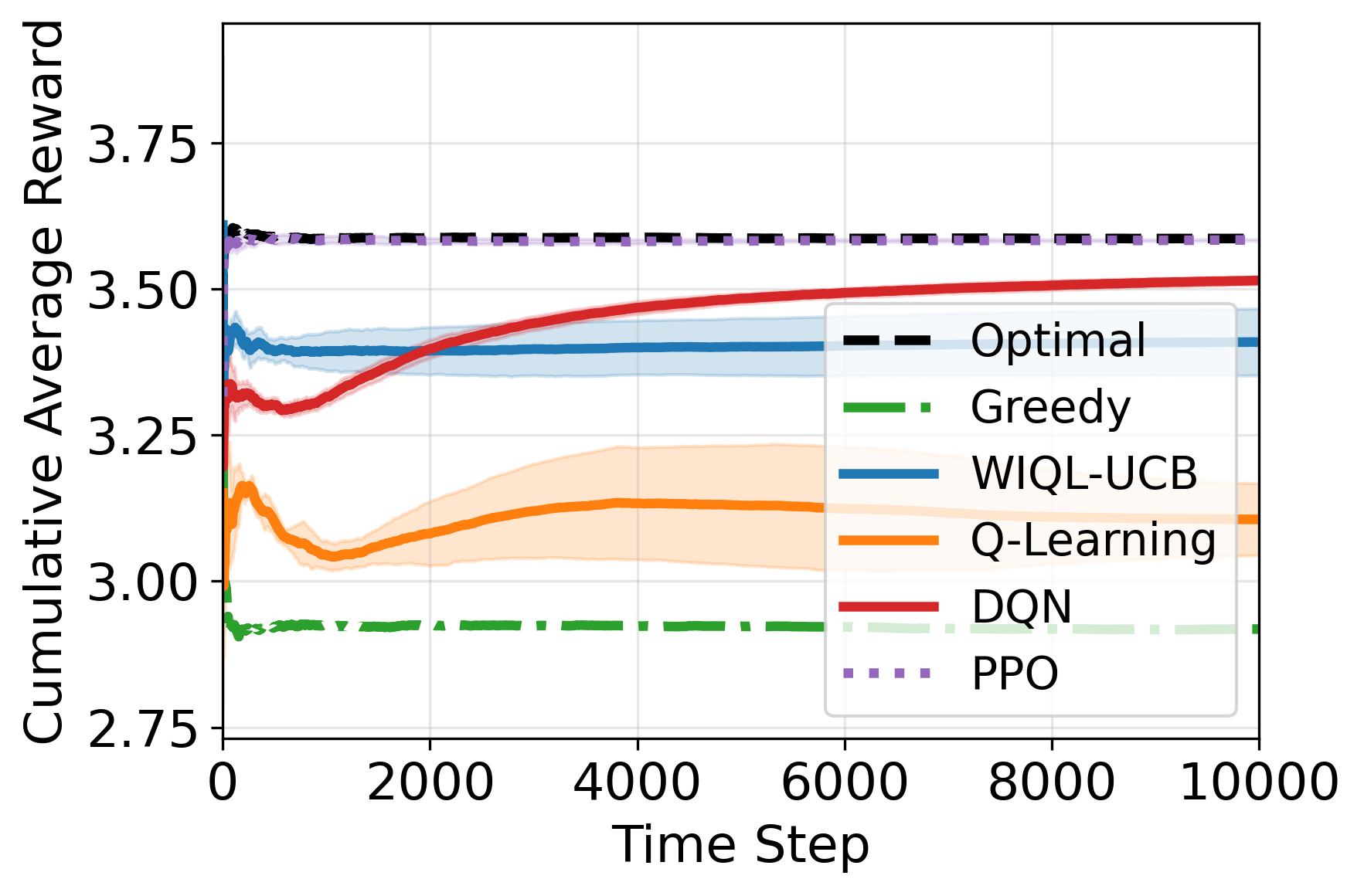}
   
  }
  \subfloat[\(N=15,\, M=3\)]{%
    \includegraphics[width=0.32\linewidth]{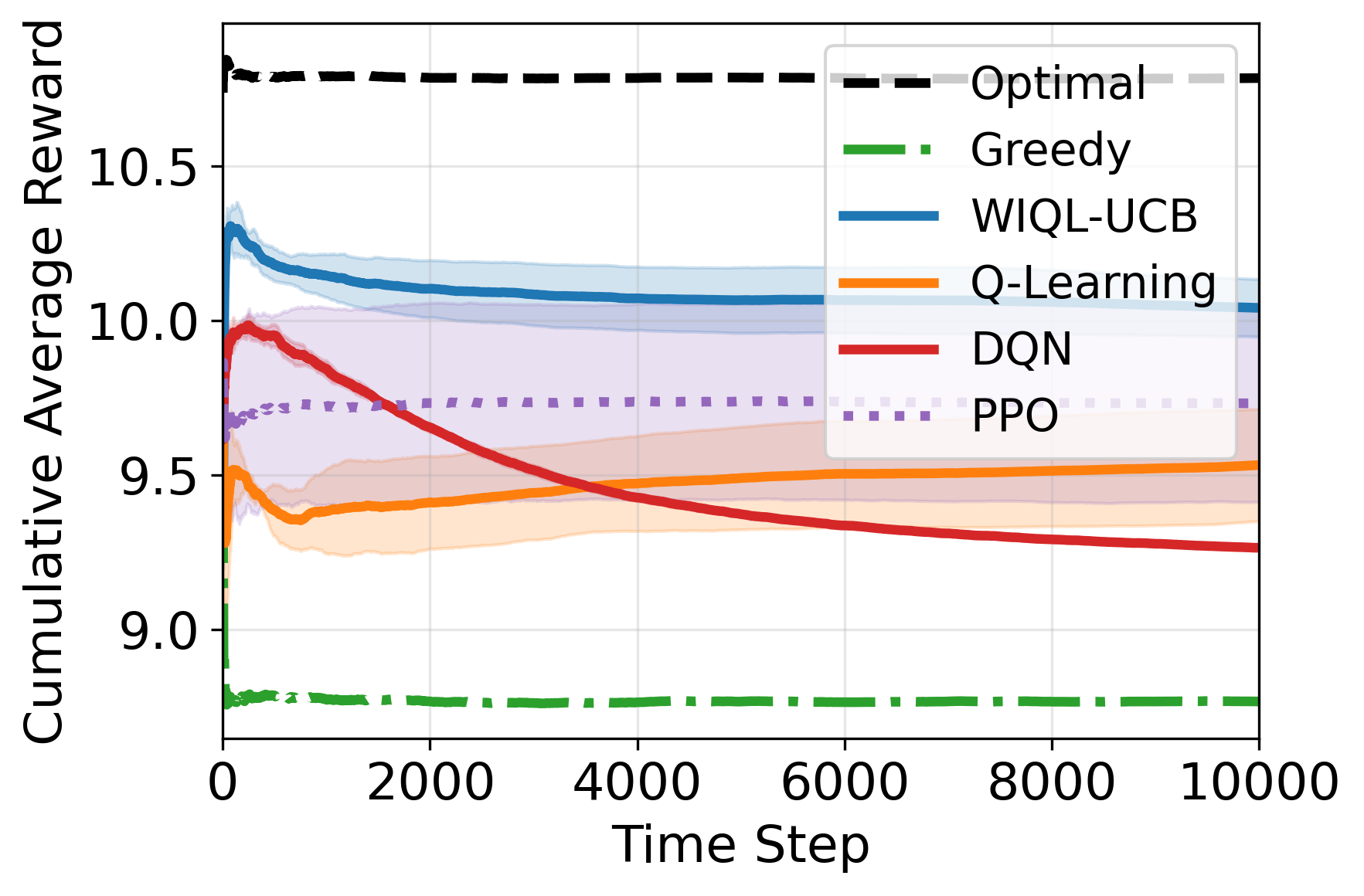}
    
  }
  \subfloat[\(N=20,\, M=4\)]{%
    \includegraphics[width=0.32\linewidth]{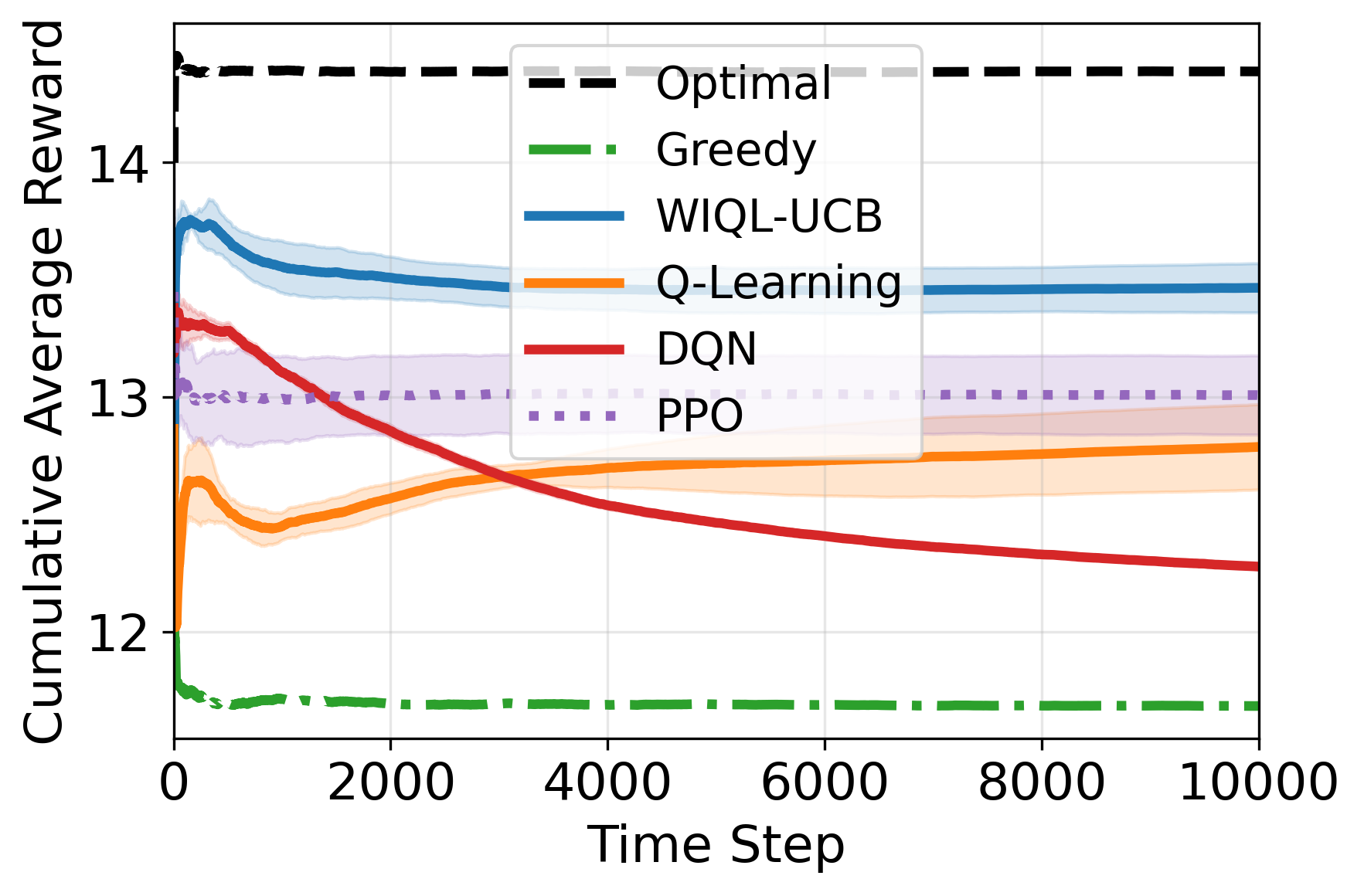}
    
  }

 \caption{Comparison of average rewards for Whittle-index and non--Whittle-index scheduling policies on the restart problem. While non--Whittle-index methods such as \gls{ppo} and \gls{dqn} perform competitively in small-scale settings, their performance deteriorates as the joint state--action space grows with increasing \(N\) and \(M\). The proposed \gls{wiql}-\gls{ucb} approach exhibits superior scalability and consistently achieves performance closest to the oracle policy.}
  \label{fig:Whittle_and_non_index_baselines_restart_prob}
\end{figure*}

\subsection{Memory Efficiency and Computational Cost of Learning-Based Scheduling Policies}
\label{sec_memory_scalability}

The memory usage and computational cost of the proposed \gls{wiql}-\gls{ucb} policy are evaluated relative to non-index learning-based baselines, including joint Q-learning, \gls{dqn}, and \gls{ppo}. To avoid ambiguity arising from scalability trends, the comparison is conducted for a fixed representative configuration with \(N = 15\) nodes and scheduling capacity \(M = 3\), as reported in Table~\ref{tab:policy_resource_comparison}.

In terms of memory usage, \gls{wiql}-\gls{ucb} exhibits a substantially smaller footprint than all joint learning approaches. The policy requires approximately 600~bytes of memory, comprising per-arm Q-values, visit counts, and Lagrange multiplier estimates. This compact representation is enabled by the index-based decomposition, which avoids maintaining a joint state--action space.

\begin{table}[h]
    \centering
    \caption{Approximate memory usage and runtime per decision for scheduling policies at \(N=15, M=3\).}
    \label{tab:policy_resource_comparison}
    \begin{tabular}{l c c}
        \hline
        \textbf{Policy} & \textbf{Memory Usage} & \textbf{Runtime (ms/step)} \\
        \hline
        \gls{wiql}-\gls{ucb} & $\approx$ 600~B & $\approx$ 0.19 \\
        Q-learning & $\approx$ 2.93~KB & 0.10 \\
        \gls{ppo} & $\approx$ 323~KB & $\approx$ 0.37 \\
        \gls{dqn} & $\approx$ 3.93~MB & $\approx$ 1.20 \\
      \hline
    \end{tabular}
\end{table}

By contrast, joint learning methods incur significantly higher memory requirements even at this moderate problem size. Joint Q-learning requires approximately 2.93~KB, corresponding to nearly a five-fold increase relative to \gls{wiql}-\gls{ucb}. Deep reinforcement learning approaches exhibit orders-of-magnitude higher memory usage: \gls{ppo} requires approximately 323~KB, while \gls{dqn} requires approximately 3.93~MB. In the case of \gls{dqn}, this overhead is dominated by the experience replay buffer, whereas for \gls{ppo} it arises from policy and value network parameters together with rollout storage. Such memory demands limit the practicality of these approaches in resource-constrained sensor network deployments.

Table~\ref{tab:policy_resource_comparison}  reports the average runtime per decision for the same configuration. Joint Q-learning achieves the lowest per-step runtime due to direct table lookup; however, this result is reported only for completeness, as joint Q-learning fails to converge reliably at larger problem sizes and is therefore not scalable in practice. Among the remaining policies, \gls{wiql}-\gls{ucb} achieves the lowest runtime per decision, requiring approximately 0.19~ms per step. The \gls{ppo} policy incurs a higher per-step cost of approximately 0.37~ms, reflecting the fixed overhead of neural network inference, while \gls{dqn} exhibits the highest runtime at approximately 1.20~ms per step the simulation parameters for the algorithms are detailed in Table~\ref{tab:algo_hyperparams}.

Taken together, these results demonstrate that \gls{wiql}-\gls{ucb} offers a favourable balance between memory efficiency and computational cost at realistic problem sizes. While neural network–based methods remain computationally feasible, their substantially higher memory usage and per-step runtime make them less suitable for deployment in large-scale or resource-limited \gls{wsns} and \gls{iot} systems. In contrast, the compact representation and low per-decision cost of \gls{wiql}-\gls{ucb} support its applicability to practical scheduling scenarios.

\section{Conclusion and Discussion}

In this work, we propose a \gls{wiql}-\gls{ucb} which does not need any hyperparamter tunning to effectively learning the optimal Whittle index of \gls{rmab} problems a problem know to be PSPACE-hard. We show via rigourous comparion that the our \gls{wiql}-\gls{ucb} outperforms other polies on benchmark examples. We then specifically focus on optimal sensor scheduling based on \gls{aoii} using the as a metric to ensure optimal scheduling for a remote monitor we adapt the edge mining technique to ensure that contrary to other techniques which assume the system dynamics at the sink is know by the node we use the edge mining technique which transmit the estimate and relevant information to allow approximation of the system state.
From the results, our proposed \gls{wiql}-\gls{ucb} effectively schedules nodes by prioritising those with faster-changing system states, thereby reducing the average Age of Incorrect Information (\gls{aoii}) while maintaining low packet transmission rates. This makes it a promising strategy for efficient scheduling in constrained \gls{wsns}, even though energy consumption was not explicitly measured in this study.
While this study focuses on sensor scheduling under the assumption of linear system dynamics where the edge mining technique is applied an important direction for future work is to extend these methods to systems with non linear dynamics.
A key advantage of the proposed \gls{wiql}-\gls{ucb} approach lies in its ability to balance exploration and exploitation without requiring manual tuning of exploration parameters. By relying on an uncertainty driven exploration mechanism, the \gls{ucb}-based method naturally prioritises actions with less certainty, which encourages more robust learning across the entire state space even in cases where certain states are rarely visited.
In contrast, Q-learning approaches that rely on fixed or adaptive \(\epsilon\)-greedy strategies suffer from a major limitation: the exploration probability \(\epsilon\) typically decays over time or adjusts based on recent performance, which can prevent the algorithm from adequately exploring infrequent or high-index states. This limitation becomes especially apparent in environments with skewed state visitation, as demonstrated in the restart example in Appendix~\ref{appendix:restart_example}, where the adaptive \(\epsilon\)-greedy policy performs poorly compared to the more consistent exploration behaviour of the \gls{ucb}-based policy.

That said, Q-learning itself is not without challenges. In environments with very large or continuous state spaces, convergence to the optimal solution remains difficult due to the sparse visitation of states and the exponential growth of the learning space Appendix~\ref{appendix_mentoring_example}. While our work shows promising results in moderate-scale problems, extending it to high-dimensional domains will require further algorithmic improvements or approximations.

\section{Appendix}

\subsection{Mentoring Instruction }
\label{appendix_mentoring_example}
We consider the \gls{rmab} mentoring example introduced by Fu \textit{et al.}~\cite{fu2019towards}. In this setting, each arm represents a student receiving mentoring support. A student can be in one of several mentoring states at any given time, and due to resource constraints, only a limited number of students can be mentored at each time step. The action \( a = 1 \) corresponds to providing mentoring, which has the potential to improve a student's state, while \( a = 0 \) corresponds to not mentoring the student.

States are indexed from 0 to 9, where higher state indices indicate better student performance. The reward function is defined as:
\[
R(s) = \sqrt{\frac{s}{10}}, \quad \text{for } s \in \{0, 1, \ldots, 9\}.
\]
This reflects diminishing marginal returns with increasing state values.
The transition probabilities for each action \( a \in \{0, 1\} \) are represented by \( 10 \times 10 \) matrices. For the active action (\(a = 1\)), a student typically improves to a higher state with probability 0.7, or regresses with probability 0.3. For the passive action (\(a = 0\)), the probabilities are reversed, indicating a higher likelihood of performance degradation when no mentoring is provided.

\textit{Active action transition matrix (\(P_1\)):}
\[
P_1 =
\begin{bmatrix}
0.3 & 0.7 & 0   & 0   & 0   & 0   & 0   & 0   & 0   & 0   \\
0.3 & 0   & 0.7 & 0   & 0   & 0   & 0   & 0   & 0   & 0   \\
0   & 0.3 & 0   & 0.7 & 0   & 0   & 0   & 0   & 0   & 0   \\
0   & 0   & 0.3 & 0   & 0.7 & 0   & 0   & 0   & 0   & 0   \\
0   & 0   & 0   & 0.3 & 0   & 0.7 & 0   & 0   & 0   & 0   \\
0   & 0   & 0   & 0   & 0.3 & 0   & 0.7 & 0   & 0   & 0   \\
0   & 0   & 0   & 0   & 0   & 0.3 & 0   & 0.7 & 0   & 0   \\
0   & 0   & 0   & 0   & 0   & 0   & 0.3 & 0   & 0.7 & 0   \\
0   & 0   & 0   & 0   & 0   & 0   & 0   & 0.3 & 0   & 0.7 \\
0   & 0   & 0   & 0   & 0   & 0   & 0   & 0   & 0.3 & 0.7 \\
\end{bmatrix}
\]

\textit{Passive action transition matrix (\(P_0\)):}
\[
P_0 =
\begin{bmatrix}
0.7 & 0.3 & 0   & 0   & 0   & 0   & 0   & 0   & 0   & 0   \\
0.7 & 0   & 0.3 & 0   & 0   & 0   & 0   & 0   & 0   & 0   \\
0   & 0.7 & 0   & 0.3 & 0   & 0   & 0   & 0   & 0   & 0   \\
0   & 0   & 0.7 & 0   & 0.3 & 0   & 0   & 0   & 0   & 0   \\
0   & 0   & 0   & 0.7 & 0   & 0.3 & 0   & 0   & 0   & 0   \\
0   & 0   & 0   & 0   & 0.7 & 0   & 0.3 & 0   & 0   & 0   \\
0   & 0   & 0   & 0   & 0   & 0.7 & 0   & 0.3 & 0   & 0   \\
0   & 0   & 0   & 0   & 0   & 0   & 0.7 & 0   & 0.3 & 0   \\
0   & 0   & 0   & 0   & 0   & 0   & 0   & 0.7 & 0   & 0.3 \\
0   & 0   & 0   & 0   & 0   & 0   & 0   & 0   & 0.7 & 0.3 \\
\end{bmatrix}
\]
\begin{figure}
    \centering
    \includegraphics[width=\linewidth]{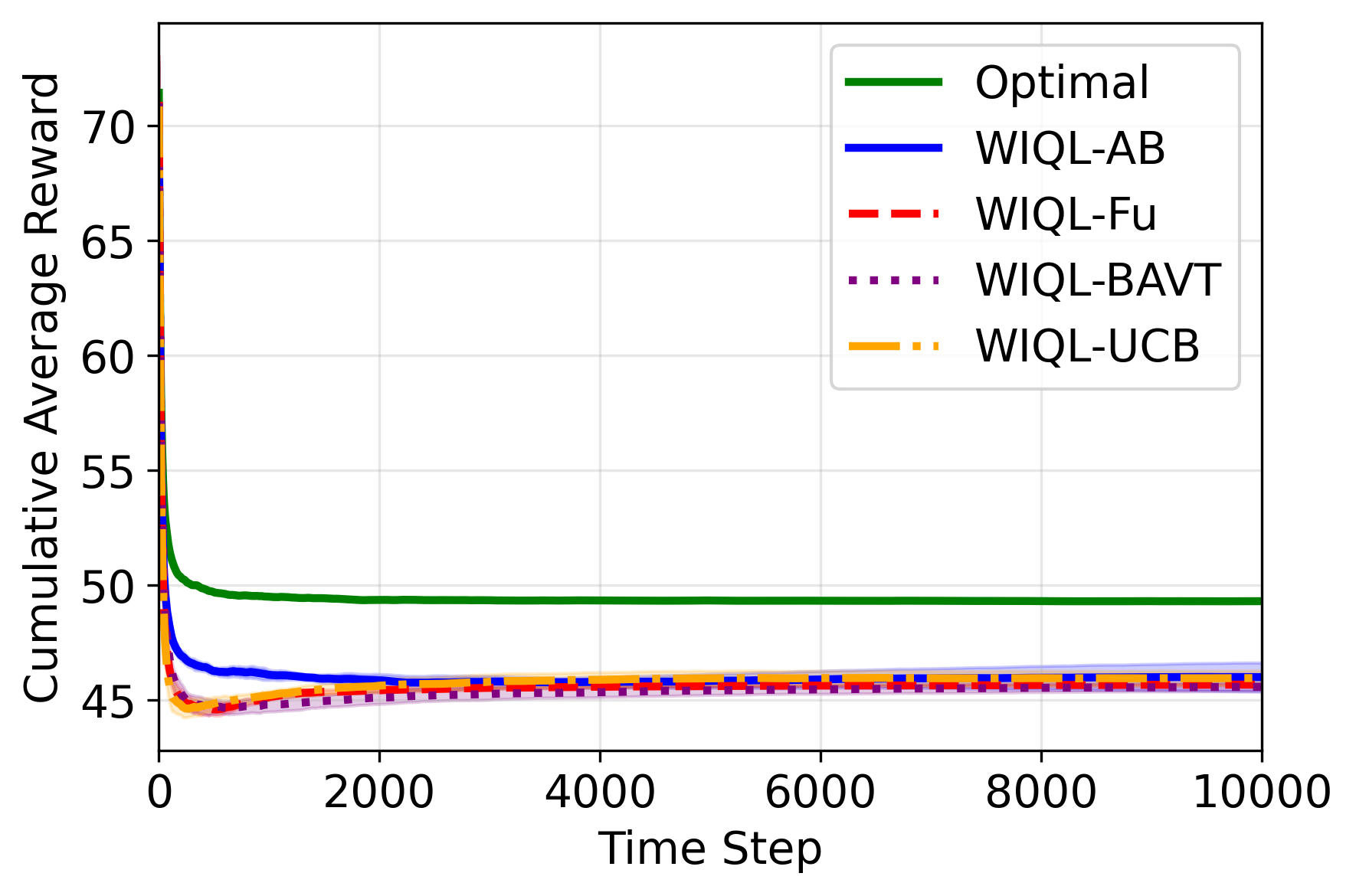}
    \caption{Mentoring example \(N=100, M=10\)}
    \label{fig_mentor_N_5_M_1}
\end{figure}
\begin{figure}
    \centering
    \includegraphics[width=\linewidth]{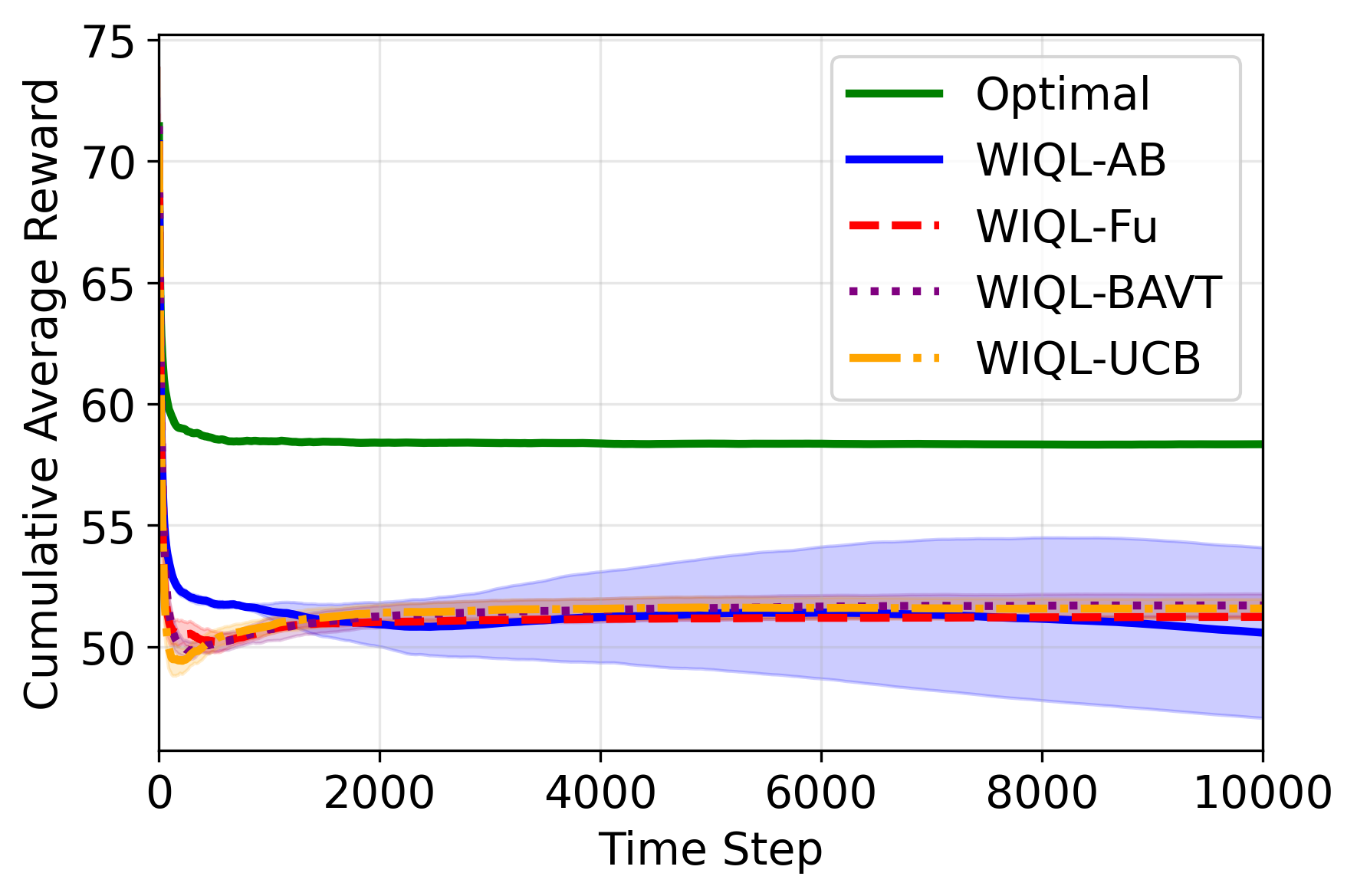}
     \caption{Mentoring example \(N=100, M=20\)}
    \label{fig:mentor_N_5_M_1}
\end{figure}
In this example, all algorithms exhibit poor performance, with wide gaps relative to the optimal solution. This reflects the inherent difficulty in achieving convergence in environments with a large state space, where certain states are rarely visited. Such sparse visitation significantly hinders the learning process and highlights a key limitation of Q-learning-based approaches in high-dimensional or under-explored settings.

\subsection{Example with restart }
\label{appendix:restart_example}
In this restart problem, the \gls{rmab} framework introduced by Avrachenkov and Borkar~\cite{avrachenkov2022whittle} is considered, where the active action forces an arm to reset to the initial state. Each arm is assumed to be in one of five states \( S = \{0, 1, 2, 3, 4\} \) at any point in time, and can take either a passive action (\( a = 0 \)) or an active action (\( a = 1 \)).

The reward function is defined as
\[
r(s, a) =
\begin{cases}
\alpha^s, & \text{if } a = 0 \quad \text{(passive)} \\
0,        & \text{if } a = 1 \quad \text{(active)}
\end{cases}
\]
where \( \alpha = 0.9 \) is a parameter controlling the exponential growth of rewards in the passive mode. Since \( \alpha < 1 \), this setup reflects diminishing returns as the state increases, thereby modelling scenarios where prolonged passive behaviour yields decreasing marginal benefit.
The state transition probabilities for each action are given as follows:
\textit{Passive mode (\(a = 0\))}: the process tends to move upward in the state space, with high probability of advancing to the next state and a small chance of remaining or regressing:
\[
P_0 =
\begin{bmatrix}
0.1 & 0.9 & 0.0 & 0.0 & 0.0 \\
0.1 & 0.0 & 0.9 & 0.0 & 0.0 \\
0.1 & 0.0 & 0.0 & 0.9 & 0.0 \\
0.1 & 0.0 & 0.0 & 0.0 & 0.9 \\
0.1 & 0.0 & 0.0 & 0.0 & 0.9 \\
\end{bmatrix}
\]

\textit{Active mode (\(a = 1\))}: taking the active action resets the system to state 0 with probability 1, regardless of the current state:
\[
P_1 =
\begin{bmatrix}
1.0 & 0.0 & 0.0 & 0.0 & 0.0 \\
1.0 & 0.0 & 0.0 & 0.0 & 0.0 \\
1.0 & 0.0 & 0.0 & 0.0 & 0.0 \\
1.0 & 0.0 & 0.0 & 0.0 & 0.0 \\
1.0 & 0.0 & 0.0 & 0.0 & 0.0 \\
\end{bmatrix}
\]

This structure models systems in which passive accumulation yields increasingly stable, though diminishing, benefits, and where occasional resets through active intervention are essential for maintaining long-term performance.
\begin{figure}
    \centering
    \includegraphics[width=\linewidth]{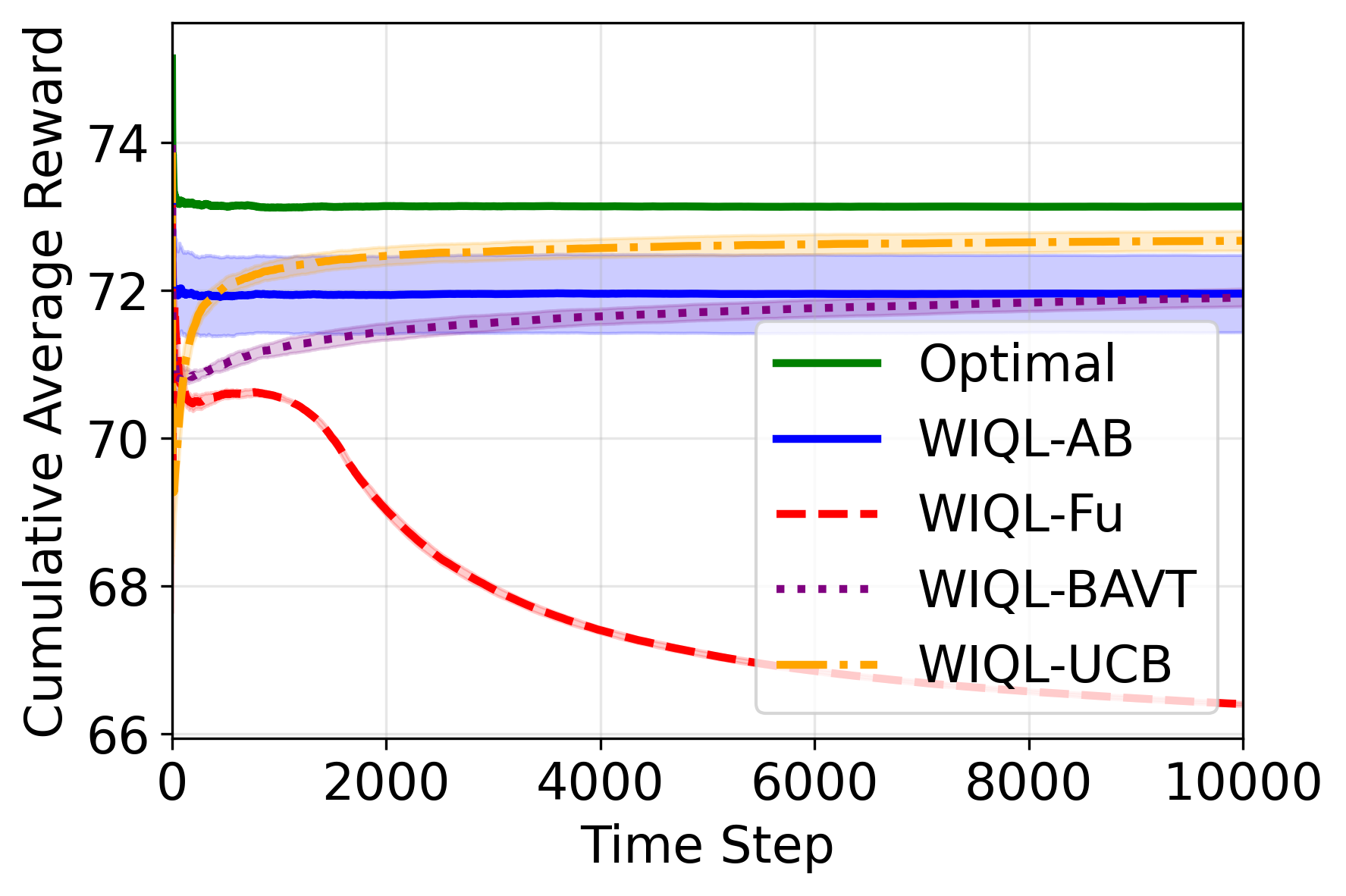}
    \caption{Restart problem \(N=100, M=10\)}
    \label{fig:restart_N_5_M_1}
\end{figure}
\begin{figure}
    \centering
    \includegraphics[width=\linewidth]{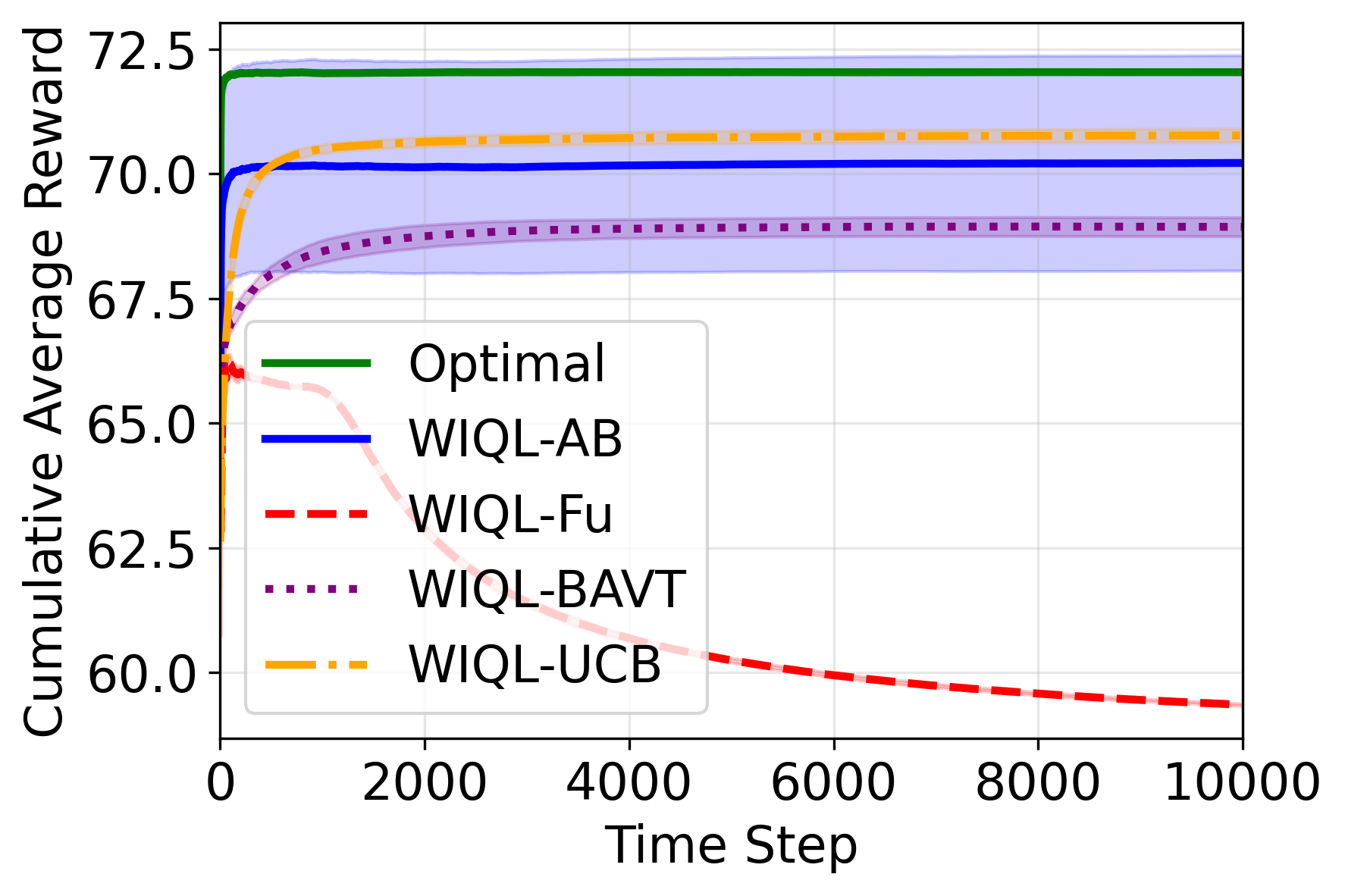}
    \caption{Restart problem \(N=100, M=20\)}
    \label{fig:restart_N_120_M_20}
\end{figure}
In this example, \gls{wiql}-BVAT performs poorly, highlighting a key limitation of the adaptive \(\epsilon\)-greedy approach. As the state space grows, higher or less frequently encountered states are visited infrequently. The adaptive \(\epsilon\) schedule reduces exploration over time without explicitly encouraging visits to rarely seen states, leading to poor coverage and suboptimal learning. In contrast, the \gls{wiql}-\gls{ucb} policy demonstrates near-optimal performance by effectively balancing exploration and exploitation, even in large or sparsely explored state spaces.

\subsection{Maternal Health Care }
\label{appendix:maternal_health}
The maternal health problem has been widely studied using the \gls{rmab} framework~\cite{wang2023optimistic,liang2024bayesian,biswas2021learn}. for this this, we consider similar example by Biswas et al.~\cite{biswas2021learn} where each arm represents a beneficiary enrolled in a maternal health programme, and weekly interventions (e.g., phone calls or health worker visits) are scheduled under resource constraints.

Biswas et al.\ model beneficiary behaviour using a three-state \gls{mdp}. At any given week, each beneficiary is assumed to be in one of three states:
\begin{itemize}
    \item  \textbf{Self-motivated \( S\):}  listens to more than 50\% of the weekly health content.
    \item  \textbf{Persuadable \( P \):}  listens to between 5\% and 50\% of the content.
    \item \textbf{Lost cause \( L\)} listens to less than 5\% of the content.
\end{itemize}

The reward function is defined based on engagement level:
\[
r(s, a) = 
\begin{cases}
0, & \text{if } s = L \\
1, & \text{if } s = P \\
2, & \text{if } s = S
\end{cases}
\]
A higher total reward thus corresponds to a larger proportion of beneficiaries in either the persuadable or self-motivated states, reflecting greater engagement with the health programme.

The simulation involves a fixed weekly intervention budget (\( M = 1000 \)) distributed among a large population (\( N = 5000 \)). Beneficiaries are divided into three categories based on their responsiveness to intervention:
\begin{itemize}
    \item \textbf{Self-motivated \( S\):} High responsiveness to intervention.
    \item \textbf{Persuadable \( P \):}  Moderate responsiveness.
    \item \textbf{Lost cause \( L\)} Minimal responsiveness.
\end{itemize}

Each category has distinct transition probabilities under passive (\(a = 0\)) and active (\(a = 1\)) actions. Below, we present the full transition matrices used for each category:

\textbf{Category A (High Improvement)}:

\textit{Passive (no intervention)}:
\[
P^{A}_0 = 
\begin{bmatrix}
0.8 & 0.2 & 0.0 \\
0.8 & 0.2 & 0.0 \\
0.0 & 0.2 & 0.8
\end{bmatrix}
\]
This matrix indicates that without intervention, beneficiaries in \(L\) and \(P\) are highly likely to remain or regress to \(L\), while those in \(S\) show strong retention.

\textit{Active (intervention)}:
\[
P^{A}_1 = 
\begin{bmatrix}
0.4 & 0.3 & 0.3 \\
0.0 & 0.2 & 0.8 \\
0.0 & 0.2 & 0.8
\end{bmatrix}
\]
This shows significant improvement with intervention, especially from \(P \rightarrow S\) with 80\% probability.

\vspace{1em}

\textbf{Category B (Moderate Improvement)}:

\textit{Passive (no intervention)}:
\[
P^{B}_0 = 
\begin{bmatrix}
0.6 & 0.4 & 0.0 \\
0.6 & 0.2 & 0.2 \\
0.2 & 0.2 & 0.6
\end{bmatrix}
\]
Moderate improvement is possible, but beneficiaries in \(P\) still risk deterioration to \(L\).

\textit{Active (intervention)}:
\[
P^{B}_1 = 
\begin{bmatrix}
0.6 & 0.2 & 0.2 \\
0.2 & 0.4 & 0.4 \\
0.1 & 0.1 & 0.8
\end{bmatrix}
\]
With intervention, beneficiaries in \(P\) now have a 40\% chance to move to \(S\).
\textbf{Category C (Low Improvement)}:

\textit{Passive (no intervention)}:
\[
P^{C}_0 = 
\begin{bmatrix}
0.6 & 0.2 & 0.2 \\
0.6 & 0.2 & 0.2 \\
0.3 & 0.3 & 0.4
\end{bmatrix}
\]
This matrix reflects limited retention and higher degradation even from \(S\).

\textit{Active (intervention)}:
\[
P^{C}_1 = 
\begin{bmatrix}
0.6 & 0.2 & 0.2 \\
0.2 & 0.6 & 0.2 \\
0.2 & 0.2 & 0.6
\end{bmatrix}
\]
Intervention has minimal effect, especially for \(P \rightarrow S\), which occurs only 20\% of the time.
These matrices collectively reflect varying degrees of beneficiary responsiveness across the three groups, with Category A being the most responsive and Category C the least. The variation highlights the importance of adaptive intervention scheduling policies.
This model captures the dynamics of real-world maternal health programmes where beneficiaries receive automated health calls. State transitions depend on both prior engagement and whether an intervention is delivered. The simulation is run over a 160-week horizon.

\begin{figure}
    \centering
    \includegraphics[width=\linewidth]{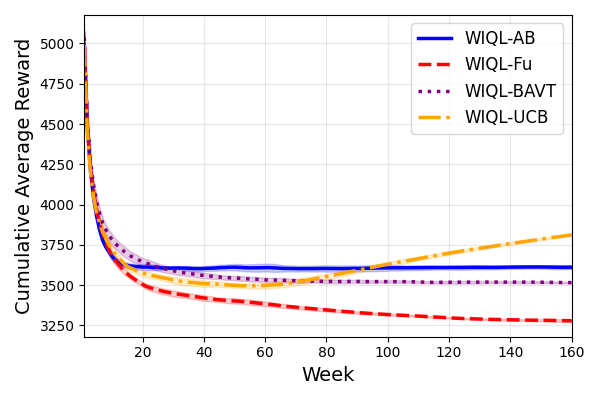}
    \caption{maternal health N=5000, M=500.}
    \label{fig:health_mdp_example_1}
\end{figure}
\begin{figure}
    \centering
    \includegraphics[width=\linewidth]{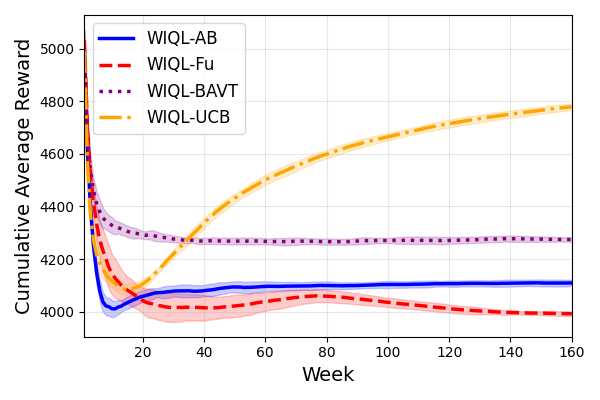}
    \caption{Maternal health, \(N = 5000\), \(M = 1500\).}
    \label{fig:health_mdp_example_2}
\end{figure}

In this example, it is observed that under higher constraint settings (i.e., smaller \(M\)), the \gls{wiql}-\gls{ucb} policy slightly outperforms other policies. However, as \(M\) increases, the performance gap becomes more pronounced, with \gls{wiql}-\gls{ucb} demonstrating significantly better results.

\begin{table}[h]
\centering
\caption{\small Algorithm-specific hyperparameters used in the numerical experiments.}
\label{tab:algo_hyperparams}
\small
\begin{tabular}{l l l}
\hline
Algorithm & Parameter & Value \\
\hline

Q-learning 
& Initial exploration probability $\epsilon_0$ 
& $0.1$ \\
& Min exploration probability $\epsilon_{\min}$ 
& $0.01$ \\
& Exploration decay factor 
& $0.95$ \\
\hline
\gls{dqn} 
& Hidden layer width 
& $64$ \\
& hidden layers
& $2$   \\
& Optimiser 
& Adam \\
& Learning rate 
& $10^{-3}$ \\
& Replay buffer size 
& $30{,}000$ \\
& Mini-batch size 
& $128$ \\
& Initial exploration probability $\epsilon_0$ 
& $1.0$ \\
& Min exploration probability $\epsilon_{\min}$ 
& $0.05$ \\
& Exploration decay factor 
& $0.999$ \\

\hline
\gls{ppo}

& Mini-batch size 
& $256$ \\
& Learning rate 
& $3 \times 10^{-4}$ \\
& Entropy coefficient 
& $0.01$ \\
& Clipping range  
& $0.2$ \\
& GAE parameter ($\lambda$) 
& $0.95$ \\
& Value function coefficient  
& $0.5$ \\
& Maximum gradient norm 
& $0.5$ \\
&  Network architecture  
& $(128,128)$ \\
& Activation function 
& Tanh \\
\hline
\end{tabular}
\end{table}

\section*{Acknowledgment}
The authors would like to thank Coventry University for the trailblazer  scholarship given to Sokipriala Jonah. This work was completed while Seong Ki Yoo was affiliated with Coventry University. BT, his current affiliation, does not express any opinion on the concepts, conclusions, or recommendations.

\bibliographystyle{IEEEtran}
\bibliography{reference} 
\end{document}

%% file: acronyms_list.tex
\newacronym{ge}{GE}{Gilbert Elliott}
\newacronym{qos}{QoS}{Quality of Service}
\newacronym{rr}{RR}{Round Robin}
\newacronym{iot}{IoT}{Internet of things}
\newacronym{wur}{WUR}{wake up radio}
\newacronym{wus}{WUS}{wake up signal}
\newacronym{pcr}{PCR}{primary communication radio}
\newacronym{ewma}{EWMA}{exponential weighted moving average}
\newacronym{ack}{ACK}{acknowledgement}
\newacronym{mse}{MSE}{mean squared error}
\newacronym{rmse}{RMSE}{root mean squared error}
\newacronym{dqn}{DQN}{Deep Q-Network}
\newacronym{ppo}{PPO}{Proximal Policy Optimisation}
\newacronym{ucb}{UCB}{upper confidence bound}
\newacronym{aoii}{AoII}{age of incorrect information}
\newacronym{aoi}{AoI}{age of  information}
\newacronym{rmab}{RMAB}{restless multi-armed bandit}
\newacronym{mdp}{MDP}{Markov decision process}
\newacronym{mab}{MAB}{multi-armed bandit}
\newacronym{pdr}{PDR}{packet delivery ratio}
\newacronym{snr}{SNR}{signal to noise ratio}
\newacronym{voi}{VoI}{value of information}
\newacronym{wsns}{WSNs}{wireless sensor networks}
\newacronym{lsip}{L-SIP}{Linear Spannish Inquisition Protocol}
\newacronym{fwaoii}{F-WAoII}{fair Whittle index AoII}
\newacronym{waoi}{W-AoI}{Whittle index AoI}
\newacronym{waoii}{W-AoII}{Whittle index AoII}
\newacronym{wiql}{WIQL}{Whittle index Q-Learning}
\newacronym{kf}{KF}{Kalman filter}
\newacronym{kfe}{KFE}{KF-Error}

%% file: glossaries_list.tex
\newglossaryentry{x}{
    name={\ensuremath{x(k)}},
    description={State representation of \ensuremath{x} at time \ensuremath{k}}
}

\newglossaryentry{xdot}{
    name={\ensuremath{\dot{x}}},
    description={Rate of change of the state \ensuremath{x}}
}

\newglossaryentry{S}{
    name={\ensuremath{\mathbf{S}}},
    description={State matrix}
}

\newglossaryentry{z}{
    name={\ensuremath{z(k)}},
    description={Measurements received by the sink at time \ensuremath{k}}
}

\newglossaryentry{H}{
    name={\ensuremath{H}},
    description={Observation matrix}
}

\newglossaryentry{alpha}{
    name={\ensuremath{\alpha}},
    description={Weighting factor in state estimation}
}

\newglossaryentry{beta}{
    name={\ensuremath{\beta}},
    description={Weighting factor for the rate of change}
}

\newglossaryentry{pdr}{
    name={PDR},
    description={Packet Delivery Ratio}
}

\newglossaryentry{snr}{
    name={SNR},
    description={Signal-to-Noise Ratio}
}